\documentclass[sigplan, 10pt]{acmart}

\acmSubmissionID{<922>}

\copyrightyear{2027}
\acmYear{2027}
\setcopyright{cc}
\setcctype{by}
\acmConference[EuroSys '27]{22nd European Conference on Computer Systems}{April 19--23, 2027}{Rabat, Morocco}
\acmBooktitle{22nd European Conference on Computer Systems (EuroSys '27), April 19--23, 2027, Rabat, Morocco}
\acmDOI{10.1145/3842654.3848573}
\acmISBN{979-8-4007-2971-3/2027/04}

\usepackage{tikz}
\usepackage{algorithm}
\usepackage{algorithmic}
\usepackage{float}
\usepackage{enumitem}
\usepackage{subcaption}
\usepackage[normalem]{ulem}
\usepackage{comment}

\newcommand{\mypar}[1]{\vspace{0.1cm} \noindent {\textbf{#1.}\/}}

\newcommand{\pjn}[1]{{\textsc{HADIS}}}

\acmBadgeR[https://www.acm.org/publications/policies/artifact-review-and-badging-current]{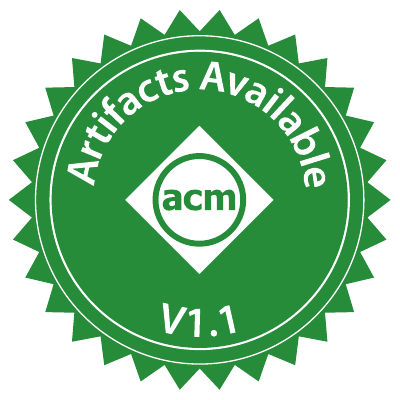}
\acmBadgeR[https://www.acm.org/publications/policies/artifact-review-and-badging-current]{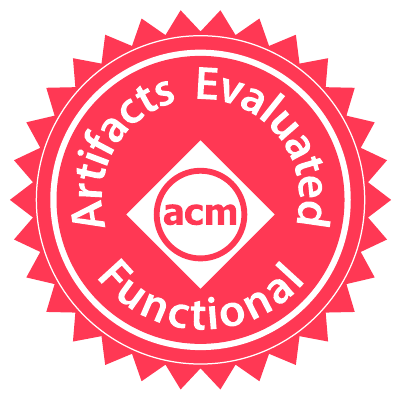}
\acmBadgeR[https://www.acm.org/publications/policies/artifact-review-and-badging-current]{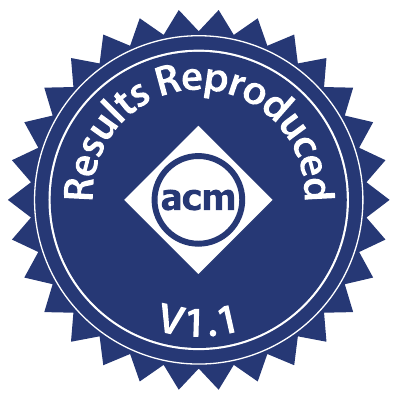}

\begin{document}

\title{HADIS: A Hybrid Architecture for Query-Aware Diffusion Model Serving}

\author{Qizheng Yang}
\affiliation{
    \institution{University of Massachusetts Amherst}
    \city{}
    \state{}
    \country{}
}
\email{qizhengyang@umass.edu}

\author{Tung-I Chen}
\affiliation{
    \institution{University of Massachusetts Amherst}
    \city{}
    \state{}
    \country{}
}
\email{tungichen@umass.edu}

\author{Siyu Zhao}
\affiliation{
    \institution{University of Massachusetts Amherst}
    \city{}
    \state{}
    \country{}
}
\email{siyuzhao@umass.edu}

\author{Ramesh K. Sitaraman}
\affiliation{
    \institution{University of Massachusetts Amherst}
    \institution{Akamai Tech}
    \city{}
    \state{}
    \country{}
}
\email{ramesh@cs.umass.edu}

\author{Hui Guan}
\affiliation{
    \institution{University of Massachusetts Amherst}
    \city{}
    \state{}
    \country{}
}
\email{huiguan@cs.umass.edu}

\begin{abstract}
Query-aware model serving routes queries through cascades of increasing cost to balance quality and throughput. Existing cascade systems force all queries through the lightweight stage, wasting resources on hard queries, and fix the model pair regardless of workload dynamics. In text-to-image diffusion serving, this waste is severe: quality is assessable only after full generation completes, and rejected lightweight outputs cannot be reused by the heavyweight model. \pjn{} is a diffusion model serving system built on a \textit{hybrid cascade architecture} that combines a pre-generation router (bypassing the lightweight model for predicted-hard queries) with a post-generation discriminator (catching false-easy cases). The hybrid architecture creates a coupled optimization space: routing thresholds, model-pair selection, and GPU allocation must be jointly adapted as workloads shift. \pjn{} reduces this space by establishing that two-model cascades match the Pareto frontier of deeper cascades, pruning candidates into a compact Pareto-optimal configuration table offline, and re-optimizing allocation at runtime via an online MILP. The hybrid design dominates both router-only and discriminator-only alternatives under mild accuracy conditions. On a 16-GPU cluster with real-world traces, \pjn{} improves response quality by up to 35\% and reduces SLO violations by 2.7--45$\times$ compared to baselines.
\end{abstract}



\maketitle

\section{Introduction}


Model serving systems must use resources efficiently to serve queries with high throughput while guaranteeing latency service-level objectives (SLOs), even as query complexity varies widely. Systems such as INFaaS~\cite{romero2021infaas}, Proteus~\cite{ahmad2024proteus}, and Cocktail~\cite{gunasekaran2022cocktail} adapt to workload dynamics by switching model variants, but treat all queries identically. \textit{Query-aware cascade serving systems}~\cite{chen2023frugalgpt, ahmad2025diffserveefficientlyservingtexttoimage, li2020cascadebert, wang2023tabi, kossmann2024cascadeserve} go further: they route each query through models of increasing cost, so easy queries exit early while hard queries proceed to expensive models. For example, DiffServe~\cite{ahmad2025diffserveefficientlyservingtexttoimage} processes every prompt with a lightweight diffusion model first; a discriminator evaluates the output and reroutes low-quality results to a heavyweight model.

Cascade serving systems face a fundamental tension. All queries must pass through the lightweight stage before quality can be assessed, wasting resources on hard queries that inevitably require the heavyweight model. Yet predicting query difficulty \textit{before} generation remains an open problem, making it risky to bypass the lightweight stage entirely. Existing systems accept this waste as unavoidable. Resolving this tension requires a serving system that coordinates pre-generation difficulty estimation with post-generation quality verification, and jointly adapts both mechanisms as query demand and difficulty composition change.


This tension is most severe in text-to-image diffusion model serving, the focus of this work.
Diffusion models such as Stable Diffusion XL~\cite{podell2023sdxl}, DALL$\cdot$E 3~\cite{dalle3}, and Diffusion Transformers~\cite{peebles2023scalablediffusionmodelstransformers} power commercial platforms including Adobe Firefly~\cite{adobefirefly} and Photoshop's generative fill~\cite{adobeps}, where users expect both visual realism and interactive responsiveness. Each image takes several seconds of GPU time, and query difficulty varies widely (from ``a banana on a white plate'' to ``a surreal cityscape representing time and memory''), making query-aware serving essential.

Two properties make cascade failures costly in this domain:
\begin{enumerate}[noitemsep, nolistsep]
    \item \textbf{No intermediate quality signal.} The denoising trajectory produces no reliable indicator of final image quality. Quality can only be assessed after full generation completes, making a post-generation check the only mechanism for detecting cascade failures.
    \item \textbf{No cross-model state transfer.} When the quality check rejects a lightweight output, the heavyweight model regenerates from scratch. Architectural heterogeneity across model families (e.g., UNet-based SDXL vs.\ DiT-based SD3.5) prevents reusing intermediate states, and rejected images are poor initializations. Every misrouted query pays the full cost of both stages.
\end{enumerate}
\noindent These properties distinguish diffusion serving from LLM serving, where speculative decoding and shared KV caches allow lightweight inference to be partially reused~\cite{leviathan2023fast}. In diffusion cascades, wasted lightweight computation is maximally costly, making it uniquely valuable to both \textit{avoid} unnecessary lightweight work and \textit{catch} unavoidable failures.

\begin{figure}[t]
    \centering
    \includegraphics[width=0.98\linewidth]{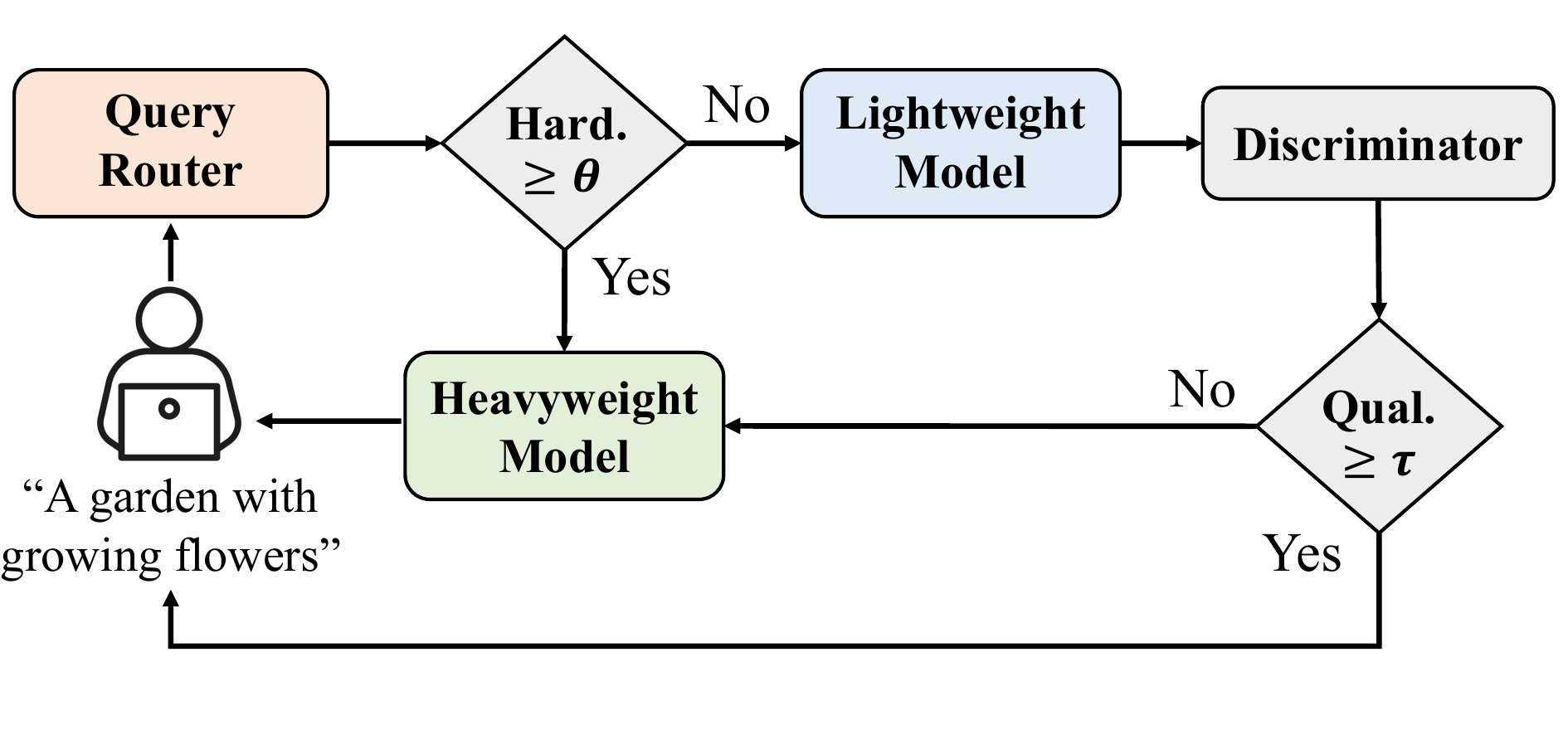}
    \vspace{-0.5cm}
    \caption{The hybrid cascade architecture in \pjn{}. The router bypasses the lightweight model for predicted-hard queries (hardness threshold $\theta$). The discriminator catches false-easy outputs and reroutes to the heavyweight model (quality threshold $\tau$). The model pair is selected at runtime.}
    \label{fig:hadis_thresholds}
\end{figure}

\vspace{.1cm}
\noindent{\bf The Hybrid Cascade Architecture.}
\pjn{} resolves this tension with a \textit{hybrid cascade architecture} (Fig.~\ref{fig:hadis_thresholds}) that combines two components: a \textit{router} that estimates query difficulty before generation and bypasses the lightweight model for predicted-hard queries, and a \textit{discriminator} that evaluates output quality after generation and catches queries the router incorrectly classified as easy. The router eliminates wasteful lightweight inference on hard queries; the discriminator prevents returning low-quality outputs on misclassified queries. Neither component alone addresses both failure modes.

\pjn{} instantiates the router as a lightweight rule-based prompt analyzer and the discriminator as a CLIP-based image evaluator. Both are pluggable: any implementation mapping queries to difficulty scores (router) or outputs to quality scores (discriminator) can be substituted.


\noindent{\bf System Design.}
The hybrid architecture introduces a coupled optimization space: cascade depth and composition, hardness and quality thresholds, and GPU allocation interact and must be jointly optimized. No existing system manages these decisions jointly. The joint space is combinatorially large (cascade configurations $\times$ threshold settings $\times$ worker allocations) and must be re-solved as query demand and difficulty composition shift, making brute-force search impractical.

\pjn{} makes this tractable with a key empirical insight: two-model cascades already capture the Pareto frontier of deeper cascades (\S\ref{sec:background}). This eliminates multi-stage configurations from both offline profiling and online optimization, reducing the full space to a compact configuration table. An online MILP re-solves over this table in milliseconds as demand fluctuates, jointly selecting the model pair, thresholds, and GPU allocation (\S\ref{sec:design}). Prior systems that fix any one dimension incur 2.7--8$\times$ higher SLO violations at comparable or worse response quality (\S\ref{sec:evaluation}).

We prove that the hybrid architecture achieves higher response quality than both router-only and discriminator-only designs at equal cost, under mild conditions on router and discriminator accuracy (\S\ref{sec:theory}). The feasibility region remains large even for moderate routing accuracy, confirming that the architecture is robust to an imperfect router.


\vspace{.1cm}
\noindent{\bf Contributions.} This paper makes four contributions:
\begin{itemize}[noitemsep, nolistsep]
    \item \pjn{}, a diffusion model serving system built on a hybrid cascade architecture that combines pre-generation routing with post-generation quality verification, dynamically selects model pairs at runtime, and jointly optimizes cascade configuration, routing thresholds, and GPU allocation via an online MILP (\S\ref{sec:design}).
    \item An empirical demonstration that two-model cascades capture the Pareto frontier of deeper cascades, enabling compact offline profiling and tractable online optimization over the coupled design space (\S\ref{sec:background}).
    \item A formal analysis deriving conditions under which the hybrid architecture dominates router-only and discriminator-only alternatives at equal cost, and showing that it is robust to imperfect routing (\S\ref{sec:theory}).
    \item An evaluation on a 16-GPU cluster with real-world traces. \pjn{} improves response quality by up to 35\% and reduces SLO violations by 2.7--45$\times$ compared to DiffServe, Proteus, INFaaS, and Clipper (\S\ref{sec:evaluation}).
\end{itemize}

\section{Background and Rationale for Design}
\label{sec:background}

This section provides background on diffusion model serving and establishes the empirical observations that motivate \pjn{}'s design: the need for dynamic model-pair selection, the sufficiency of two-model cascades, and the value of hybrid routing. 
All measurements in this section use the first 5K caption-image pairs of \textit{ImageNet-1K}~\cite{imagenet15russakovsky}, with captions as prompts and images as the reference set.

\subsection{Background}

\begin{figure}[t]
    \centering
    \includegraphics[width=\linewidth]{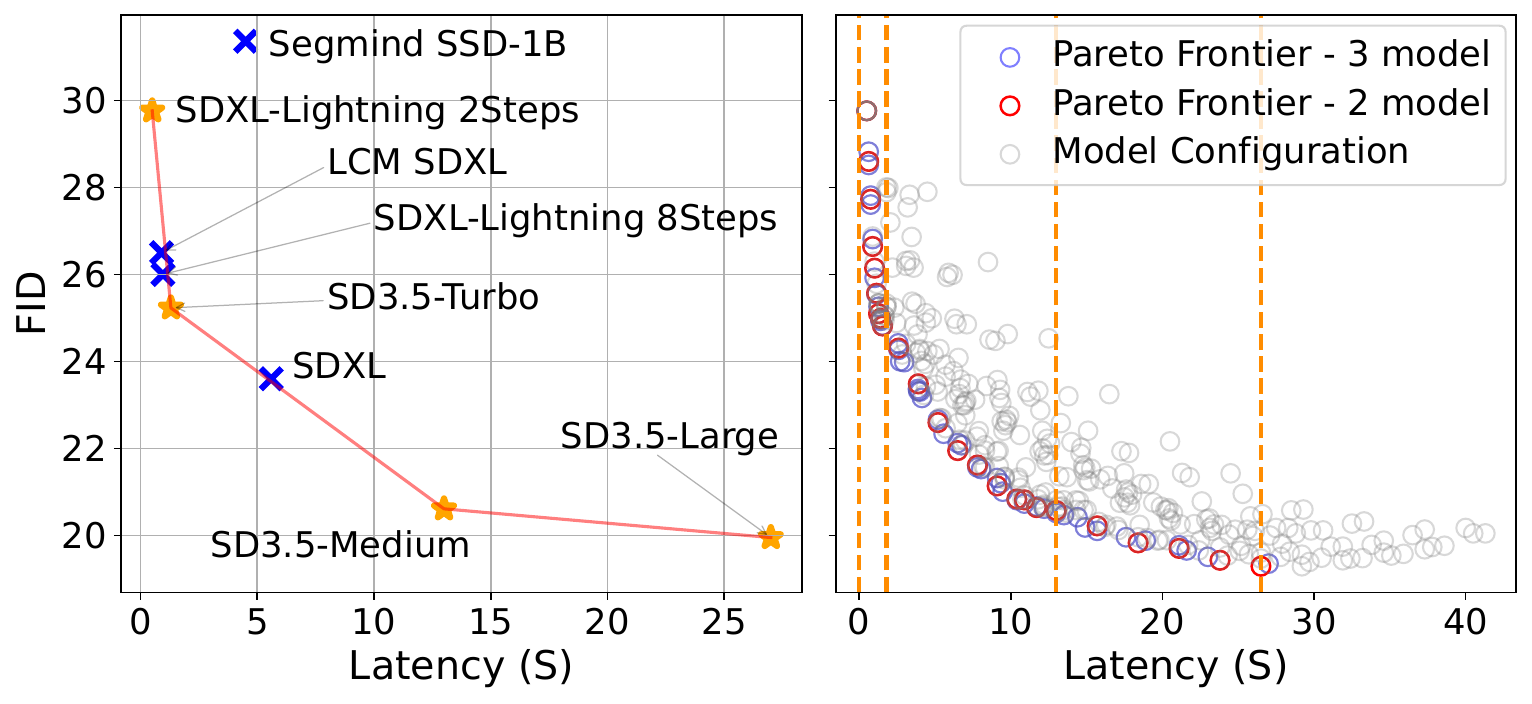}
    \vspace{-0.6cm}
    \caption{Quality-latency tradeoffs of different diffusion model variants \textit{(left)} and model cascades \textit{(right)}. Quality is measured by Fréchet Inception Distance~\cite{10.5555/3295222.3295408} (FID, lower is better). The cascades are constructed by the model variants colored in orange in the left panel. The orange dashed lines partition latency into three regimes where each regime has a distinct model cascade that yields the frontier points. }
    \label{fig:motivation_quality_latency}
\end{figure}

\mypar{Background for T2I diffusion models}
Text-to-image (T2I) diffusion models exhibit a clear quality-latency trade-off (Figure~\ref{fig:motivation_quality_latency}-\textit{left}): more compute (deeper networks, more denoising steps) yields better image quality but longer runtime and lower throughput. In our setting on L40S GPUs, the spectrum spans sub-second generation (e.g., SDXL-Lightning at 2 steps $\approx$ 0.5s) to tens of seconds (e.g., SD3.5-Large at 50 steps $\approx$ 27s) for an image of 1024$\times$1024 resolution.

\mypar{Why diffusion makes cascade failures costly}
Two properties make cascade failures uniquely expensive in diffusion serving. Evidence from recent systems supports each.

For the first property (\textit{no intermediate quality signal}), the denoising trajectory yields no reliable mid-generation quality indicator. Recent adaptive-step methods~\cite{zhang2025adadiff, wang2024atedm} reduce per-step computation or select which steps to execute, but their decisions are based on per-step difficulty or attention redundancy, not on whether the \textit{final} output will satisfy a quality threshold. Failures can only be detected by a post-generation discriminator after the full lightweight inference completes.

For the second property (\textit{no cross-model state transfer}), architectural heterogeneity across model families prevents reusing intermediate states across variants. For example, UNet-based models (SDXL) use a 4-channel VAE latent space, while DiT-based models (SD3, SD3.5~\cite{esser2024scaling}) use a 16-channel latent space with a different noise schedule (rectified flow vs.\ DDPM). These differences make latent states dimensionally and semantically incompatible across families. Even within a single architecture, partial-denoising methods like SDEdit~\cite{meng2022sdedit} require running around 80\% of the denoising steps to produce acceptable results from an existing image, offering minimal savings over generation from scratch. Modern T2I serving systems reflect this constraint: NIRVANA~\cite{295597} caches intermediate denoising states only within a single model variant, and MoDM~\cite{modm2026} reuses cached images only with smaller models, generating from scratch when invoking the heavyweight model to preserve quality. A rejected lightweight output is therefore a sunk cost, not a usable starting point for the heavyweight model.

These properties distinguish diffusion from LLM serving, where speculative decoding~\cite{leviathan2023fast} reuses lightweight tokens via cheap token-level verification and KV cache reuse.

\subsection{Rationale for Design}

\mypar{Rationale for dynamic model choice}
No fixed lightweight/heavyweight pair remains optimal across varying query demands and latency budgets.

Figure~\ref{fig:motivation_quality_latency}-\textit{right} shows that the best cascade configuration varies with the latency budget. The orange dashed lines partition latency into three regimes. Within each regime, the frontier points arise from a different cascade: <SDXL-Lightning, SD3.5-Turbo> at low latency, <SD3.5-Turbo, SD3.5-Medium> at mid latency, and <SD3.5-Medium, SD3.5-Large> at high latency. When the SLO is tighter, a faster cascade is preferred; when the SLO relaxes, higher-quality cascades can be chosen.

\begin{figure}[t]
    \centering
    \includegraphics[width=\linewidth]{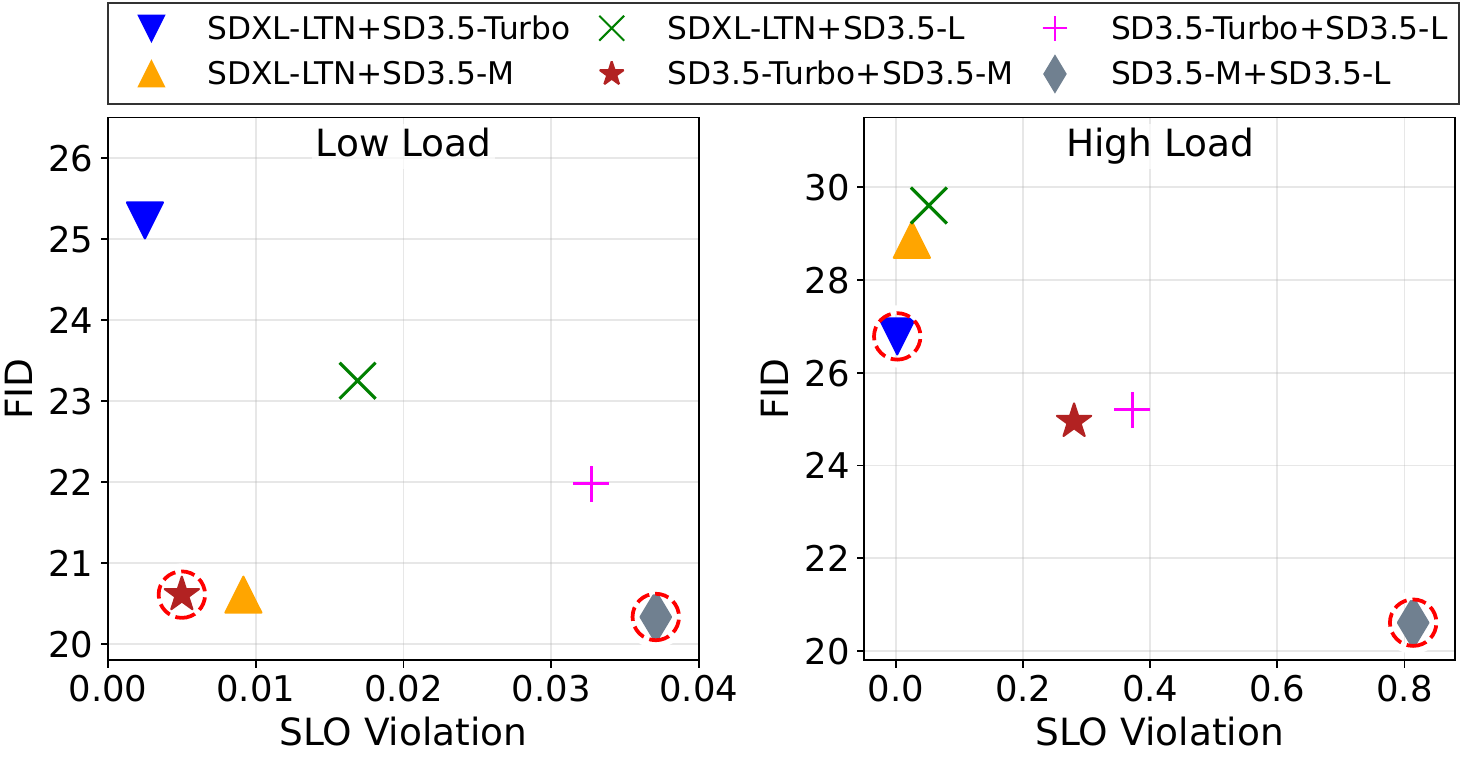}
    \vspace{-0.5cm}
    \caption{Limitations of fixed model cascades. In both panels, each point represents a distinct two-model cascade. Comparing between low load (\textit{left}) and high load (\textit{right}), the best cascade (circled in red) varies as the system workload and the tolerable SLO violation ratio change. }
    \label{fig:motivation_model_combination}
\end{figure}

Figure~\ref{fig:motivation_model_combination} shows the best cascade also varies with query load, comparing several two-model cascades under low demand (left) and high demand (right) with the SLO held fixed. Each cascade uses the resource allocation algorithm from DiffServe~\cite{ahmad2025diffserveefficientlyservingtexttoimage}. At low load, all pairs keep SLO violations small. <SD3.5-Turbo, SD3.5-Medium> (red star) attains the best violation rate with competitive FID, while <SD3.5-Medium, SD3.5-Large> trades a slight increase in violations for the best FID (grey diamond). Under high load, pairs involving slower models cannot meet throughput demands and incur large violations, whereas <SDXL-Lightning, SD3.5-Turbo> (blue triangle) remains within SLO and delivers the best quality-latency balance.

The optimal pair varies with both the latency budget and the workload, because different pairs occupy different segments of the quality-latency Pareto frontier and saturate at different throughputs. This motivates dynamic model-pair selection at runtime, but also expands the optimization space that the system must manage.

\mypar{Rationale for two-model cascades}
Dynamic model-pair selection raises a combinatorial challenge: with $n$ candidate models, $n(n-1)$ ordered pairs exist, each with its own threshold settings. We show that restricting cascades to two stages incurs no quality loss, substantially shrinking this space.

Figure~\ref{fig:motivation_quality_latency}-\textit{right} explores the design space using four model variants (orange stars in the left panel). For each candidate cascade we enumerate routing policies specified by the fraction of queries routed to the heavyweight model(s), and measure the resulting mean latency (x-axis) and FID (y-axis) on 5k prompts. The right panel plots all feasible cascade configurations in grey by varying model combinations and routing policies. The Pareto frontiers\footnote{A Pareto frontier $P$ is a set of model cascades such that for every $m \in P$ there is no other model cascade $m' \in P$ that has {\it both} lower FID and latency than $m$.} obtained by two-model and three-model cascades are colored in red and blue, respectively.
Two facts stand out. First, the red and blue frontiers coincide over the relevant latency range: for any latency budget, a two-model cascade attains essentially the same best FID as a three-model cascade. Second, while the blue frontier fills in a few intermediate points, it does not deliver better quality at a given latency.

Adding a third stage offers little benefit but introduces system overhead: an extra first-stage inference pass, additional discriminator evaluations, increased queuing between stages, and more routing/placement decisions. Two-model cascades already realize the full Pareto trade-off, so \pjn{} adopts a two-stage design. This reduces the cascade configuration space from exponential (all subsets of models) to quadratic ($n(n-1)$ pairs), and with candidate pruning (\S\ref{sec:adaptive_cascade}), further down to a handful of profiled configurations.

\mypar{Rationale for hybrid architecture}
Neither a router nor a discriminator alone suffices. We quantify the cost of each single-component design.

A router-only design requires the router to be both lightweight and highly accurate. However, estimating query hardness for text-to-image generation remains an open problem~\cite{10.1007/978-3-642-12275-0_52, 10.1145/3715275.3732158, 10.1145/3491102.3501825}, making high-accuracy routing difficult without expensive learned models.

A discriminator-only design forces all queries through the lightweight model regardless of difficulty. When the lightweight model is itself expensive, this overhead is substantial. For instance, with the pair <SD3.5-Medium, SD3.5-Large> (latencies $\sim$13s and $\sim$27s), the mandatory lightweight pass adds 33\% to end-to-end latency for hard queries that inevitably require the heavyweight model.

\pjn{} combines both: the router sends obviously hard queries directly to the heavyweight model, and the discriminator catches false-easy cases after lightweight inference. Our evaluation (\S\ref{sec:ablation_routing}) confirms that hybrid routing reduces SLO violations versus discriminator-only and improves FID versus router-only.

\section{\pjn{} Design} \label{sec:design}

\pjn{} is a diffusion model serving system motivated by the observations in \S\ref{sec:background}. This section describes its architecture (\S\ref{sec:overview}), adaptive model cascading (\S\ref{sec:adaptive_cascade}), hybrid routing (\S\ref{sec:hybrid_routing}), and resource management (\S\ref{sec:resource_management}).

\subsection{Overview of \pjn{}} \label{sec:overview}

\begin{figure}[t]
    \centering
    \includegraphics[width=\linewidth]
    {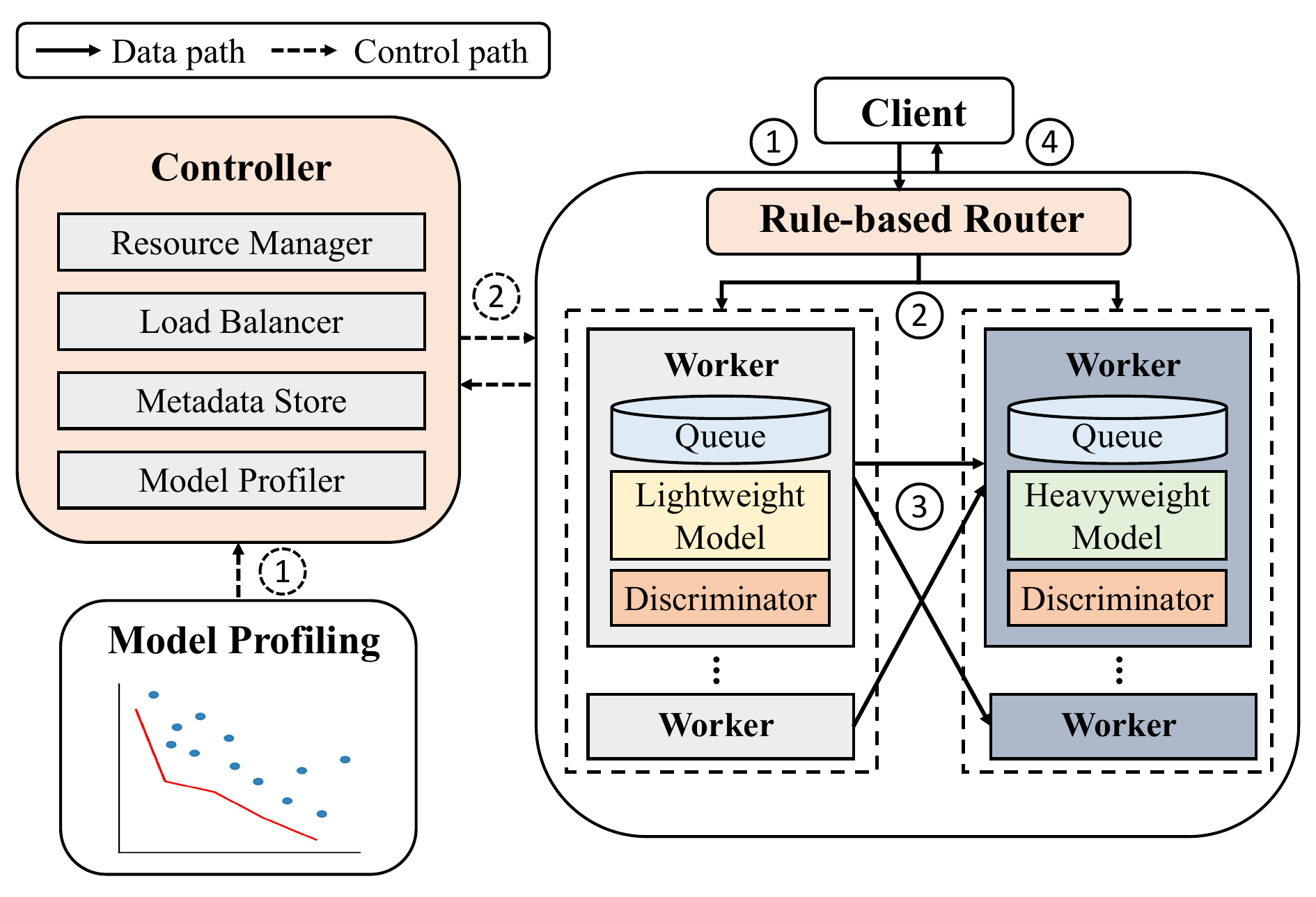}
    \vspace{-0.7cm}
    \caption{System Architecture of \pjn{}.} 
    \vspace{-0.2cm}
    \label{fig:system_architecture}
\end{figure}

Fig.~\ref{fig:system_architecture} shows the system architecture of \pjn{}. It has three major components: Controller, Query Router, and Workers.

\textit{Control path} (dotted arrows in Fig.~\ref{fig:system_architecture}). \pjn{} is initialized with a set of diffusion models registered in the Controller. The control path takes two actions: (1) The Controller profiles registered models and pairwise combinations offline to generate a cascade configuration lookup table. (2) At runtime, the Controller leverages this table to select the optimal cascade configuration, including a model pair and the associated thresholds.

\textit{Data path} (solid arrows in Fig.~\ref{fig:system_architecture}). (1) A query from the client is sent to the query router. (2) The router estimates query hardness and routes the query to a worker hosting a suitable model variant. (3) If the query is processed by the lightweight model and the quality score falls below the threshold, the worker re-routes the query to a heavyweight model. (4) The final output is returned to the client.

We next describe each system component.

\textbf{Controller.}
A centralized Controller manages resources and coordinates system modules. It comprises: 
(1) A \textit{Resource Manager} that periodically solves MILP to determine the model pair and thresholds (i.e., the router's \textit{hardness threshold} and the discriminator's \textit{quality threshold}) and to produce a resource allocation plan (i.e., model placement across workers and batch sizes);
(2) A \textit{Load Balancer} that generates routing tables so each query is directed to a worker, and updates these tables when model selection or placement changes.  
(3) A \textit{Metadata store} that periodically pulls heartbeats from the Query Router and Workers to update execution statistics, including per-model queue lengths and demand at each worker. 
(4) A \textit{Model Profiler} that supplies a lookup table of profiled cascade configurations to the Resource Manager. Only registered diffusion-model pairs and discriminators are eligible for adaptive model cascading.
\S\ref{sec:adaptive_cascade} describes construction of the lookup table and \S\ref{sec:resource_management} presents the resource-management formulation based on the lookup table.

\textbf{Workers.}
Each worker hosts either a lightweight model (co-located with a discriminator) or a heavyweight model, maintaining a local queue and processing its queries. Roles are swapped at runtime to handle time-varying demand. 
The Controller determines the batch size, model variant to host, and the quality threshold for workers, while workers report status (e.g., queries received, queue lengths, etc) for planning.
A worker that receives a new assignment finishes the request it is executing, then offloads its current model and loads the new one concurrently. Requests already in its queue wait for the duration of the transition. Each worker keeps exactly one model resident, and the cost is measured in \S\ref{sec:overhead_analysis}.

\textbf{Query Router}
The router sits on the data path between clients and workers. It scores query hardness and assigns obviously hard queries directly to heavyweight models, sending the rest to lightweight models (\S\ref{sec:hybrid_routing}).

\textbf{Shared Serving}
All users share one deployed cascade and one worker pool rather than a dedicated cascade per user. Requests from different users interleave in the same per-model queues, so the Controller plans on aggregate demand and shared queue lengths. The MILP charges both queueing and execution latency against the SLO (\S\ref{sec:resource_management}), so a rising request rate drives the next planning cycle toward faster configurations and shifts workers accordingly. \pjn{} does not isolate resource per user or enforce fairness among users.

\subsection{Adaptive Model Cascading} \label{sec:adaptive_cascade}
Since two-model cascades capture the Pareto frontier of deeper cascades (\S\ref{sec:background}), \pjn{} restricts the cascade to a single lightweight/heavyweight pair and dynamically selects the pair at runtime. However, determining the optimal pair and resource-allocation plan is NP-hard: the search space spans model candidates, decision thresholds, model placements, and batch-size choices. \pjn{} decomposes this into two stages: offline model profiling (this section) and online MILP-based resource management (\S\ref{sec:resource_management}). 

\mypar{Model candidate selection}
Exhaustively profiling every possible cascade is unnecessary. \pjn{} prunes the model pool with three rules before offline benchmarking: 
(1) If multiple models exhibit \textit{similar} latency and FID, we retain a single representative and discard the others; 
(2) If one model \textit{Pareto-dominates} another (no worse in latency and strictly better in quality, or vice versa), we keep the dominating model and drop the dominated ones;
(3) If a model lies approximately on the line segment between two neighbors, we omit it, as a two-stage cascade can reproduce its trade-off by thresholding between the neighbors.

Figure~\ref{fig:motivation_quality_latency}-\textit{left} illustrates the retained (orange) and pruned (blue) models. In our running example, we start with 8 model candidates, yielding 28 two-model cascades and 2800 configurations (10 hardness thresholds and 10 quality thresholds per cascade). After pruning 4 models, only 4 remain, yielding 6 cascades and 600 configurations, reducing profiling cost by 79\%. On our L40S GPU, the full profiling completes in approximately 78 GPU-hours (\(\sim\)58 GPU-hours for one-time generation on the evaluation set, plus \(\sim\)20 GPU-hours for FID computation across the configurations).
Image generation is done once per model and grows linearly with the pool size, while FID computation grows quadratically with it. A larger pool also enlarges the configuration tables the MILP searches (\S\ref{sec:overhead_analysis}), but it does not raise the cost of serving a query, since the cascade always holds two models and each worker hosts one of them.

\mypar{Offline profiling and lookup-table construction}
For each two-model cascade <lightweight, heavyweight>, \pjn{} sweeps the hardness threshold and quality threshold over a fixed grid. Each sweep point defines a cascade configuration recorded as:
\begin{itemize}[leftmargin=*, nolistsep, noitemsep]
    \item \textit{model\_pairs: } the pair instantiating the lightweight and heavyweight roles
    \item \textit{diff\_thres:} the router's hardness threshold \textcolor{teal}{($\theta$)}
    \item \textit{qual\_thres:} the discriminator's quality threshold \textcolor{teal}{($\tau$)}
    \item \textit{route\_ratio: } the measured fraction of queries processed by each model under these thresholds \textcolor{teal}{($r_0$ and $r_1$)}
    \item \textit{FID: } the resulting response quality
\end{itemize}
We then prune dominated rows to retain only Pareto-optimal configurations. The union over all pairs yields a compact lookup table of <pair, thresholds> configurations that captures the relevant quality–latency trade-offs and induced routing fractions. This table is the sole input to the online phase, where the Resource Manager jointly selects a configuration, model placement, and batch sizes via a fast MILP, enabling \pjn{} to adapt model pairs and thresholds at runtime without expensive online exploration.
Route ratios are measured during profiling, so the table carries a ratio for every configuration, including configurations that are not currently deployed.

\subsection{Hybrid Routing} \label{sec:hybrid_routing}
The hybrid architecture requires two components: a difficulty scorer that maps a query to a hardness score, and a quality scorer that maps a generated output to a quality score. These are \textit{pluggable} interfaces; any conforming implementation can be substituted. \pjn{} instantiates the difficulty scorer as a rule-based router that estimates query hardness \textit{before} generation, and the quality scorer as a CLIP-based discriminator that evaluates generated images. We evaluate this instantiation against non-hybrid alternatives in \S\ref{sec:ablation_routing}.

\mypar{Rule-based router}
The Query Router implements a lightweight rule-based prompt analyzer that scores query hardness from textual features including prompt length, token rarity, object count, abstractness, attribute density, spatial relations, action verbs, and named entities. Features are extracted, normalized into a compact vector, and combined via a weighted sum with weights tuned by grid search to optimize the quality-latency objective.

\mypar{CLIP-based discriminator}
The Discriminator is co-located with lightweight models on GPUs and determines whether their outputs are aesthetically acceptable. Accepted results are returned to the client, while rejected ones are re-routed to heavyweight models. We use a CLIP backbone~\cite{radford2021learningtransferablevisualmodels} because it provides robust image embeddings for aesthetic evaluation~\cite{10.3389/frai.2022.976235}. The discriminator keeps the pretrained CLIP frozen and fine-tunes a multi-layer perceptron (MLP) head.

\textit{Training.} 
We train the discriminator as a binary classifier to distinguish high-quality, real images (e.g., ImageNet~\cite{imagenet15russakovsky}) from images generated by our candidate diffusion models. This teaches it to detect typical divergences, such as loss of sharpness, texture incoherence, and structural glitches, enabling accurate quality assessment within the cascade.

\textit{Inference.} 
Given a prompt and its generated image, CLIP produces text and visual embeddings; the MLP head consumes the visual embedding and outputs a softmax score in $[0, 1]$, used as the \textit{quality score} (likelihood the image is ``real-like'' and thus high quality). The overhead of the discriminator is analyzed in \S\ref{sec:overhead_analysis}.

\subsection{Resource Management} \label{sec:resource_management}
The Resource Manager adapts \pjn{} to the current workload by jointly \textit{(i)} selecting the cascade configuration (model pair and thresholds), \textit{(ii)} placing models across workers, and \textit{(iii)} choosing batch sizes per model. Under low demand, the Resource Manager favors higher-quality configurations (lower hardness threshold, higher quality threshold); under high demand, it shifts toward faster configurations to protect SLO compliance. These changes are coordinated with model placement and batch-size updates to process the demand within the latency SLO.

\mypar{The resource management problem}
We formulate the problem with cascade configuration selection as a mixed-integer linear program (MILP) over a pre-profiled lookup table of cascade configurations (described in \S\ref{sec:adaptive_cascade}). 
At runtime, the Resource Manager selects one configuration and decides model placement and batch sizes per model so that demand is satisfied and the latency SLO is met while minimizing FID (maximizing response quality). 

\textit{Optimization variables.} For each model pair, we introduce a binary selector $z_k$ over the threshold configuration $k$ of that pair, where $k=\langle\tau, \theta\rangle$ encodes a hardness threshold and a quality threshold. For each model $i$ in the pair, we choose an integer worker count $x_i \ge 0$ and a one-hot batch size via binaries $y_{i,b}$ over a discrete set of batches $b$.

\textit{Inputs from profiling.} The lookup table provides the routing fraction $r_{k,i}$ to model $i$ under configuration $k$, the profiled latency $L_{i,b}$ and throughput $\mu_{i,b}$ at batch size $b$, and the response quality FID$_k$. Let $\lambda$ denote the system demand, $S$ the available workers, and $T_{SLO}$ the latency deadline. Queueing delay for model $i$ is estimated online from the instantaneous queue length $Q_i$ and the arrival rate $\lambda_i$.

\textit{Constraints.} The variables must satisfy three groups of constraints. First, considering the resource availability, each model selects a single batch size, each cascade has exactly one threshold configuration for the pair, and worker usage is bounded by the budget:
\begin{equation}
    \sum_{b} y_{i,b}=1, \quad 
    \sum_{k} z_k=1, \quad  
    \sum_{i} x_i \leq S \quad  \forall i
\end{equation}
Second, considering the throughput constraint, per-model capacity must cover the share of demand routed to that model under the chosen configuration. We apply an over-provisioning factor $\beta$ to the demand $\lambda$:
\begin{equation}
    x_i \sum_{b} y_{i,b}\mu_{i,b} \ge \beta\lambda \sum_{k} z_k r_{k,i} \quad\quad \forall i
\end{equation}
Setting $\beta > 1$ allocates against an inflated demand and prioritizes SLO compliance at some quality cost, while $\beta < 1$ frees capacity for heavier cascades that improve quality at the risk of more violations. $\beta$ is a deployment-time knob that traces the quality-SLO trade-off curve characterized in \S\ref{sec:static_traces}.
Then, considering the latency constraint, end-to-end latency must remain within the SLO. Let the selected batch latency for model $i$ be $\widehat{L}_i = \sum_b y_{i,b}L_{i,b}$ and let $D^{\text{queue}}_i = \alpha \frac{Q_i}{\lambda_i}$ denote the queueing delay estimate given Little's law~\cite{shortle2018fundamentals} where $\alpha$ is a safety factor. We impose a conservative path bound, making sure the latency constraint is met:
\begin{equation}
    \sum_{i}(\widehat{L}_i + D^{\text{queue}}_i) \leq T_{SLO}
\end{equation}
Eventually, the objective quality variable equals the FID of the selected configuration:
\begin{equation}
    F = \sum_k z_k \mathrm{FID}_k
\end{equation}

\textit{MILP formulation.} The resource manager identifies the optimal configuration and allocation by
\begin{equation*}
\min_{\,\{z_k\},\,\{x_i\},\,\{y_{i,b}\}} \;  F
\quad \text{s.t. Eqs.~(1)--(4)} 
\end{equation*}

\mypar{Solving the MILP}
The optimization problem is solved periodically by invoking an MILP solver to identify a global optimal cascade configuration, model allocation, and model batch sizes. We estimate query demand $\lambda$ using an exponentially weighted moving average (EWMA) on the recent demand history. 
\pjn{} therefore plans against observed traffic, not oracle knowledge of future arrivals.
Note that the time overhead to solve the MILP does not lie on the critical path of query serving as the MILP is called asynchronously. We provide a detailed analysis of the overhead in \S\ref{sec:overhead_analysis}.

\vspace{-0.2cm}
\section{Formal Analysis of Hybrid Architecture} \label{sec:theory}

\begin{figure}[t]
    \centering
    \includegraphics[width=0.49\linewidth]{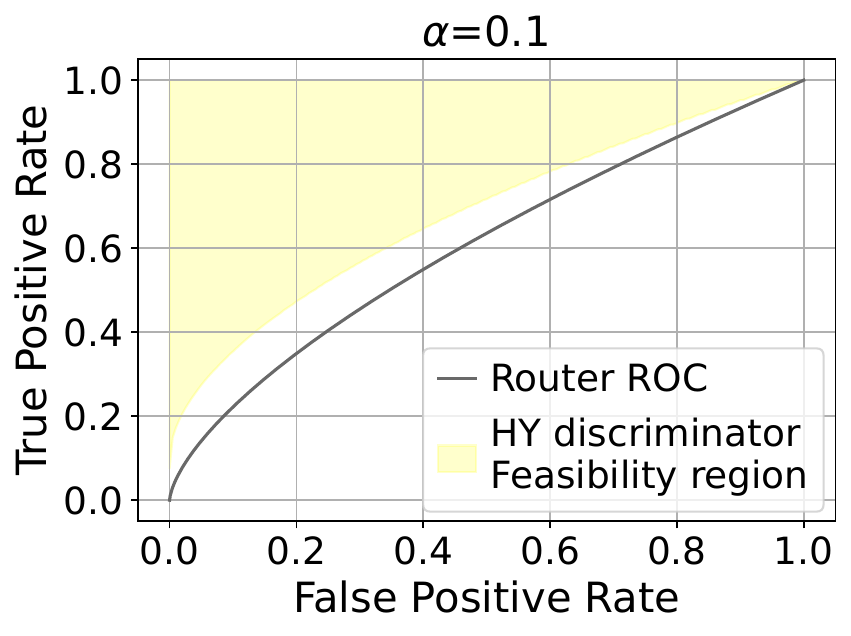}
    \includegraphics[width=0.49\linewidth]{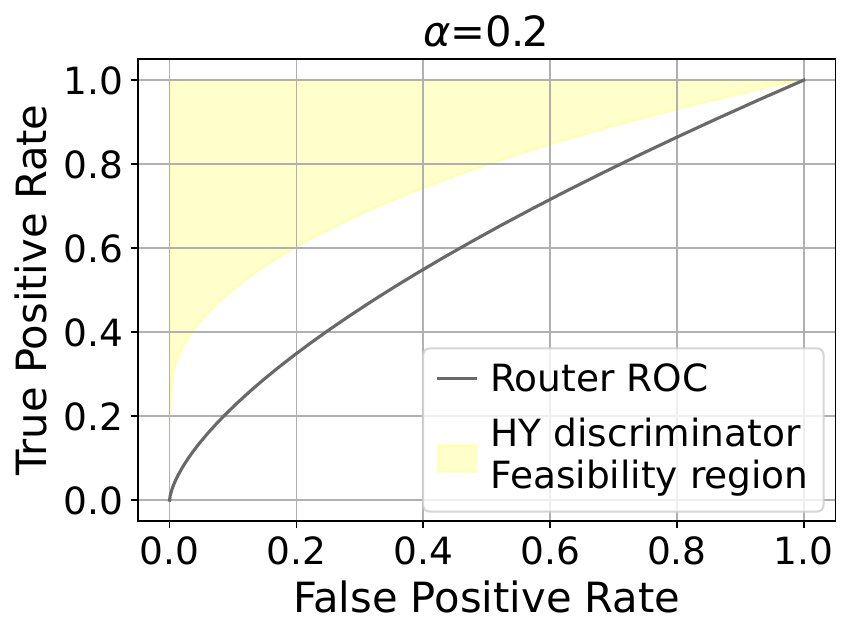}
    \caption{Feasibility region for the discriminator of HY architecture so that HY outperforms RO. Higher $\alpha$ means smaller gaps between lightweight and heavyweight models. }
    \label{fig:HY_vs_RO}
\end{figure}

\begin{figure}[t]
    \centering
    \includegraphics[width=0.49\linewidth]{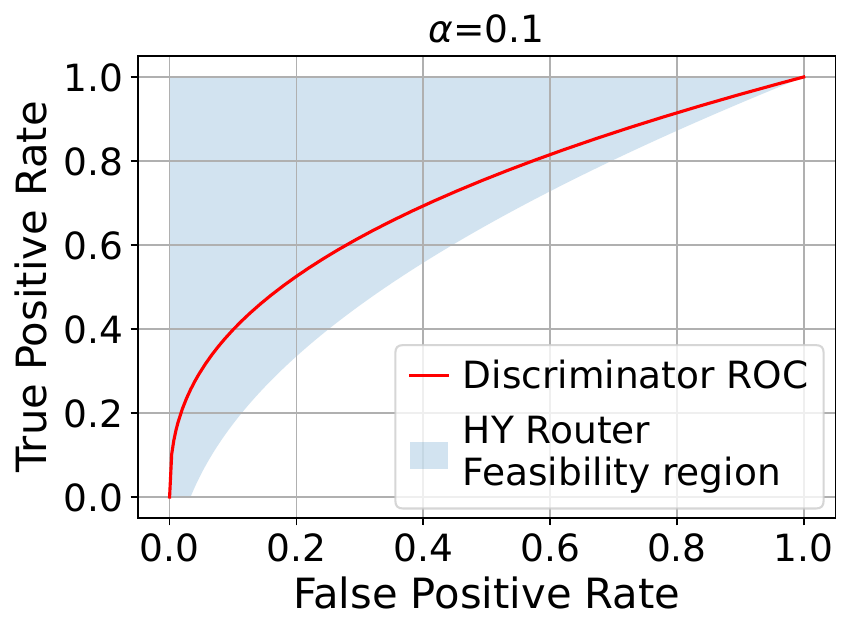}
    \includegraphics[width=0.49\linewidth]{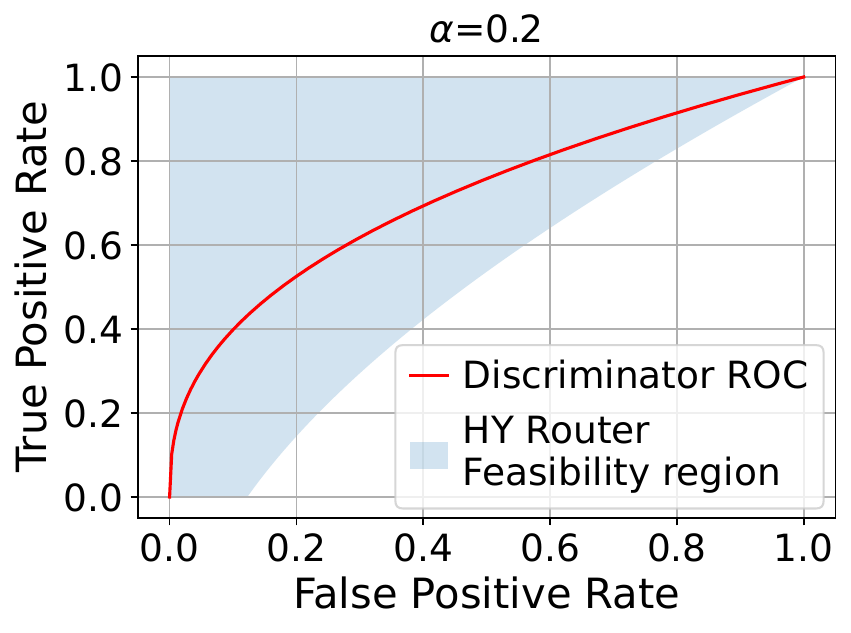}
    \caption{Feasibility region of the router of HY architecture so that HY outperforms DO. As $\alpha$ increases, the region becomes larger. }
    \label{fig:HY_vs_DO}
\end{figure}

This section formally validates the hybrid architecture presented in \S\ref{sec:overview}. We show that under mild conditions, the hybrid architecture (\textbf{HY}) achieves better quality than Router-Only (\textbf{RO}) and Discriminator-Only (\textbf{DO}) at equal expected cost. We interpret these conditions geometrically as \emph{feasibility regions} in the ROC planes of the router and the discriminator (Fig.~\ref{fig:HY_vs_RO} and Fig.~\ref{fig:HY_vs_DO}).

\mypar{Preliminary}
The analysis considers a hybrid architecture with four components: a cheap, fast model $M_0$, a slower, high-quality model $M_1$, a router, and a discriminator. The router decides whether each query is sent to $M_0$ or directly to $M_1$, while the discriminator decides, for queries that went through $M_0$, which outputs are good enough to return and which should be escalated to $M_1$, as illustrated in Fig.~\ref{fig:hadis_thresholds}. 
Let $q$ denote the fraction of easy queries in the workload, so hard queries occupy mass $1-q$. Let $s_0$ and $s_1$ be the latencies of running $M_0$ and $M_1$ with $s_0 < s_1$. We define the \textbf{latency ratio} $\alpha=\frac{s_0}{s_1}\in(0,1)$ which captures how much cheaper the lightweight model is relative to the heavyweight model. 

For the router with threshold $\theta$, we define $r_{TP}^{\theta}$ as the probability an easy query is routed to $M_0$ (i.e., True Positive Rate, TPR), and $r_{FP}^{\theta}$ as the probability a hard query is routed to $M_0$ (i.e., False Positive Rate, FPR). As $\theta$ varies, the pair $(r_{FP}^{\theta}, r_{TP}^{\theta})$ traces the router's ROC curve. Similarly, for the discriminator with threshold $\tau$, we define $d_{TP}^\tau$ as the probability that an output from $M_0$ on an easy query is accepted (TPR), whereas $d_{FP}^\tau$ is the probability that an output on a hard query is incorrectly accepted and completed at $M_0$ (FPR). As $\tau$ varies, the pair $(d_{FP}^{\tau}, d_{TP}^{\tau})$ traces the discriminator's ROC curve. Together, these two ROC curves describe the operating space in which we will derive conditions for HY to dominate RO and DO under equal expected cost.

To reason about quality analytically, we define quality using a simple probabilistic model. We assume the heavyweight model $M_1$ always produces acceptable quality for both easy and hard queries, while the lightweight model $M_0$ is adequate only for easy queries. Under this abstraction, the only way to harm quality is to return an $M_0$ output on a hard query. The \textbf{quality} ($Q$) is hence defined as one minus the probability of this failure event. We model \textbf{cost} ($C$) as the expected latency for an architecture to process a query.

\mypar{Comparison between HY and RO}
In a RO architecture, a hard query is mishandled if and only if the router classifies it as easy and sends it to $M_0$, which occurs with probability $(1-q)\,r_{FP}^{\theta_R}$, yielding:
\begin{equation*}
\begin{aligned}
    {\text{Quality:}\quad} Q_{RO} &= 1 - (1-q)r_{FP}^{\theta_R}, \\
    {\text{Cost:} \quad } C_{RO} &= r_0^{\theta_R}s_0 + r_1^{\theta_R}s_1, \\
    s.t. \quad r_0^{\theta_R} &= qr_{TP}^{\theta_R}+(1-q)r_{FP}^{\theta_R}, \quad r_1^{\theta_R}=1-r_0^{\theta_R},
\end{aligned}
\end{equation*}
where $\theta_R$ is the router threshold of RO architecture; $r_0$ and $r_1$ are the fractions of queries routed to $M_0$ and $M_1$, respectively.

In a HY architecture, a hard query is mishandled only if it is routed to $M_0$ \textbf{and} the discriminator incorrectly accepts the $M_0$ output. This happens with probability $(1-q)\,r_{FP}^{\theta_H} d_{FP}^{\tau_H}$:
\begin{equation*}
\begin{aligned}
    {\text{Quality:}\quad} Q_{HY} &= 1-(1-q)r_{FP}^{\theta_H}d_{FP}^{\tau_H}, \\
    {\text{Cost:} \quad } C_{HY} &= r_0^{\theta_H}s_0 + (r_1^{\theta_H}+\pi^{\theta_H,\tau_H})s_1, \\
    s.t. \quad r_0^{\theta_H} &= qr_{TP}^{\theta_H}+(1-q)r_{FP}^{\theta_H}, \quad r_1^{\theta_H}=1-r_0^{\theta_H}, \\
    \pi^{\theta_H,\tau_H} &= qr_{TP}^{\theta_H} (1-d_{TP}^{\tau_H}) + (1-q)r_{FP}^{\theta_H}(1-d_{FP}^{\tau_H}),
\end{aligned}
\end{equation*}
where $\pi$ is the fraction of queries that are first routed to $M_0$ and then rerouted to $M_1$.

To compare HY against RO, we impose the equal-cost constraint $C_{HY} = C_{RO}$ and analyze the quality gap. 
Note that only if $\theta_H>\theta_R$ can the cost be equal.
Requiring $Q_{HY} \ge Q_{RO}$ (i.e., HY is at least as good as RO at the same cost) yields the following bounds on the discriminator ROC point $(d_{FP}^{\tau_H}, d_{TP}^{\tau_H})$ of the HY architecture:
\begin{equation}\label{eq:HY_vs_RO}
\begin{aligned}
    d_{TP}^{\tau_H} \geq &\frac{\alpha\left(q(r_{TP}^{\theta_H}-r_{TP}^{\theta_R}) + (1-q)(r_{FP}^{\theta_H}-r_{FP}^{\theta_R})\right)+qr_{TP}^{\theta_R}}{qr_{TP}^{\theta_H}}, \\
    d_{FP}^{\tau_H} \leq & \frac{r_{FP}^{\theta_R}}{r_{FP}^{\theta_H}}.
\end{aligned}
\end{equation}

The inequalities define a \textit{feasibility region} in the ROC plane for configuring $\tau_H$ : \textbf{as long as HY employs a discriminator whose $(\text{FPR},\text{TPR})$ lies inside this region, the HY architecture will match RO's expected cost while achieving better response quality.} We visualize such a region in Fig.~\ref{fig:HY_vs_RO} by fixing $q=0.5$ and traversing the feasible $(r^\theta_{TP}, r^\theta_{FP})$ pairs. 
As $\alpha$ increases from $0.1$ to $0.2$, the feasible region shrinks. This is because when the lightweight model becomes more expensive relative to $M_1$, only discriminator thresholds with sufficiently high $d_{TP}$ and low $d_{FP}$ (i.e., making accurate re-routing decisions) allow HY to still dominate RO at equal cost. In practice, the discriminator ROC lies well above the router ROC, making these conditions practically attainable. 
Therefore, the hybrid architecture tolerates an imperfect router: a modestly accurate discriminator anywhere in this region compensates for routing mistakes by catching low-quality $M_0$ outputs.

\mypar{Comparison between HY and DO}
We now compare the hybrid architecture against a DO architecture, where every query first runs through $M_0$ and a discriminator with threshold $\tau_D$ decides whether to accept the $M_0$ output or reroute the query to $M_1$. 
\begin{equation*}
\begin{aligned}
    {\text{Quality:}\quad} Q_{DO} &= 1-(1-q)d_{FP}^{\tau_D}, \\
    {\text{Cost:}\quad} C_{DO} &= s_0 + \pi^{\tau_D}s_1, \\
    s.t.\quad
    \pi^{\tau_D} &= q(1-d_{TP}^{\tau_D}) + (1-q)(1-d_{FP}^{\tau_D}).
\end{aligned}
\end{equation*}

Requiring $Q_{HY} \ge Q_{DO}$ and $C_{HY}=C_{DO}$ yields bounds on the router ROC point of our hybrid architecture (i.e., $\langle r_{FP}^{\theta_H}, r_{TP}^{\theta_H}\rangle$) for a given discriminator ROC and latency ratio $\alpha$. With $d_{TP}^{\tau_H} > \alpha$, we obtain
\begin{equation}\label{eq:HY_vs_DO}
\begin{aligned}
    r_{TP}^{\theta_H} \geq \frac{\alpha(1-q)r_{FP}^{\theta_H} + qd_{TP}^{\tau_D} - \alpha}{q(d_{TP}^{\tau_H}-\alpha)}, \quad
    r_{FP}^{\theta_H} \leq \frac{d_{FP}^{\tau_D}}{d_{FP}^{\tau_H}}
\end{aligned}
\end{equation}

The above inequalities define a \textit{feasibility region} in the router ROC plane: \textbf{if HY employs a router whose $(\text{FPR},\text{TPR})$ lies inside this region, HY is guaranteed to match DO's expected cost while achieving better response quality.}
Fig.~\ref{fig:HY_vs_DO} illustrates how the required router accuracy depends on $\alpha$. When the lightweight model is very cheap ($\alpha=0.1$), DO can already exploit $M_0$ aggressively at low cost, so the feasibility region is relatively small and the router must be quite accurate to provide additional benefit. As $\alpha$ increases, the region expands, indicating that even a moderately accurate router is sufficient for HY to dominate DO. 
The router bypasses the lightweight stage for clearly hard queries, avoiding the $M_0$ inference cost that DO would pay on every query.

In summary, the feasibility regions for both comparisons are practically attainable: the discriminator easily outperforms the router in classification accuracy (making HY$>$RO feasible), and even a simple rule-based router provides sufficient accuracy when the lightweight model is non-trivial (making HY$>$DO feasible). \S\ref{sec:ablation_routing} confirms this empirically.

\section{Evaluation} \label{sec:evaluation}
This section evaluates \pjn{}. We outline the experimental setup (\S\ref{sec:exp_settings}), then report end-to-end performance on real-world (\S\ref{sec:real_traces}) and static (\S\ref{sec:static_traces}) workloads. 
Ablation studies quantify the contributions of routing and model-pair selection (\S\ref{sec:ablation_routing}) and the MILP-based resource allocator (\S\ref{sec:ablation_milp}). We then analyze runtime adaptation on synthetic traces (\S\ref{sec:synthetic_traces}), study sensitivity to latency SLOs (\S\ref{sec:ablation_slo}) and query hardness (\S\ref{sec:eval_hardness}), and report system overheads (\S\ref{sec:overhead_analysis}).

\subsection{Experiment settings}\label{sec:exp_settings}

\textbf{Implementation of \pjn{}.}
We implement \pjn{} in $\sim$5K lines of Python code, using HuggingFace~\cite{von-platen-etal-2022-diffusers} and PyTorch~\cite{paszke2019pytorchimperativestylehighperformance} to execute the diffusion models and discriminator, spaCy~\cite{honnibal2020spacy} to extract prompt features for the rule-based router, and Gurobi~\cite{gurobi} to solve the MILP optimization. Our testbed cluster consists of 16 workers, each with an NVIDIA L40S GPU, and system components (Controller, Load Balancer, Worker, Client) communicate via gRPC for low-latency coordination. 
We assume a fixed GPU allocation throughout each experiment, which is standard in cost-budgeted serving deployments and matches the evaluation setup of DiffServe and Proteus. 
Autoscaling the cluster size is orthogonal to \pjn{}'s design and can be layered on top.

\textbf{Model Selection.}
We apply our model-selection rules to choose four diffusion models with diverse latency-quality trade-offs, configured at their default step counts and image resolution to ensure high-quality generation. SDXL-Lightning~\cite{lin2024sdxllightningprogressiveadversarialdiffusion} is a fast generative model that produces an image in 2 steps, yielding $\sim$0.5s latency per image on an L40S GPU. Stable Diffusion 3.5~\cite{sd35} is a diffusion transformer family that can produce images of high quality. We configure SD3.5-Turbo with 4 steps ($\sim$1.3s per image on L40S), while SD3.5-Medium and SD3.5-Large use 50 steps, with latencies of $\sim$13s and $\sim$27s, respectively, on the same hardware. Unless otherwise noted, the latency SLO is 60s, chosen as a multiple of the slowest model's runtime.

\textbf{Datasets and workloads.}
For quality evaluation, we use \textit{ImageNet-1k}, a widely-used subset of ImageNet~\cite{imagenet15russakovsky}, with associated captions. We take the first 5k caption-image pairs, using captions as prompts and the corresponding images as the reference set for FID computation. 
To train the discriminator, we use a disjoint set of 10k ImageNet images and 10k images generated by both lightweight and heavyweight models. Training takes $\sim$790s on one L40S GPU.
To drive system load, we use the Microsoft Azure Functions trace~\cite{shahrad2020serverless} as a representative real-world workload, applying shape-preserving scaling to match our cluster's capacity. 
\S\ref{sec:real_traces}, \S\ref{sec:ablation_routing}, \S\ref{sec:ablation_milp} and \S\ref{sec:ablation_slo} all run on this trace.

\textbf{Evaluation Metrics.} We use two primary metrics. (1) \textit{Response quality FID} measures the distance between the distribution of generated images and a reference set of real images (lower is better). For each system configuration, we compute FID across all queries against the references of our evaluation set. (2) \textit{SLO violation ratio} is the fraction of queries that exceed the deadline in end-to-end latency or are proactively dropped because they are predicted to miss it. 

\begin{figure}[t]
    \centering
    \includegraphics[width=\linewidth]{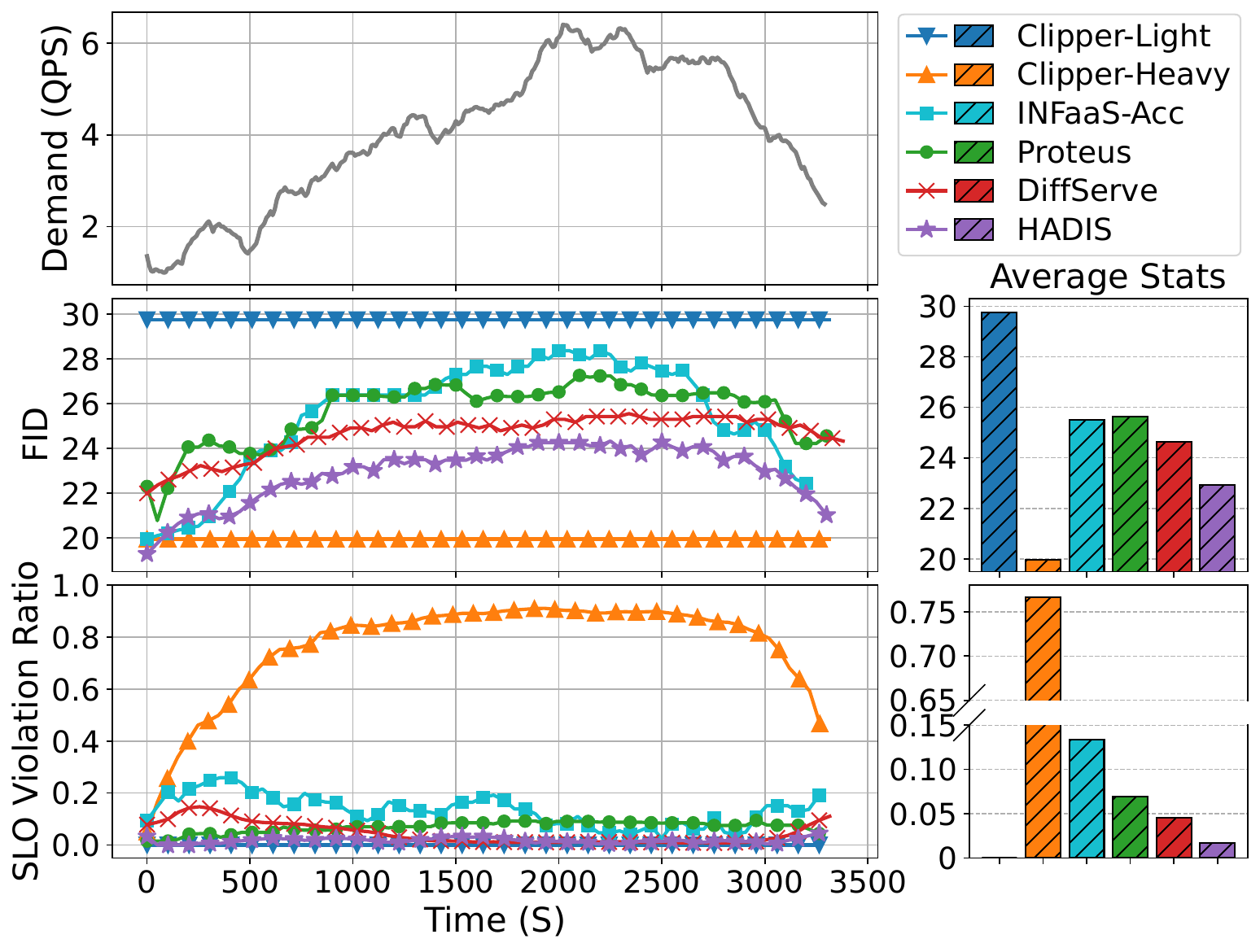}
    \vspace{-0.7cm}
    \caption{End-to-end comparison on real-world traces.
    On average, \pjn{} reduces FID by 7-23\% over all baselines except Clipper-Heavy, and reduces SLO violation by 2.7$\times$, 4.2$\times$, 8$\times$, and 45$\times$ compared to DiffServe, Proteus, INFaaS-Acc, and Clipper-Heavy. Clipper-Light's SLO violation is 0.}
    \label{fig:end2end_results}
\end{figure}

\textbf{Baselines.}
We compare \pjn{} against static serving (Clipper), dynamic variant selection (INFaaS, Proteus), and query-aware cascading (DiffServe).
\begin{itemize}[leftmargin=*, nolistsep, noitemsep]
    \item \textbf{Clipper}~\cite{crankshaw2017clipper} is a static baseline with no dynamic model selection or query-aware routing. We implement two variants. \textbf{Clipper-Light} minimizes SLO violations with the lightest model (SDXL-Lightning), and \textbf{Clipper-Heavy} maximizes response quality with the heaviest model (SD3.5-Large). Clipper also represents other static serving systems (e.g., TensorFlow-Serving~\cite{tfserving}) that rely on developers for resource allocation.
    \item \textbf{INFaaS}~\cite{romero2021infaas} dynamically selects model variants to satisfy a user objective. Its original formulation minimizes cost subject to a quality constraint, the inverse of our setting. We therefore evaluate \textbf{INFaaS-Acc}, the adaptation used in Proteus~\cite{ahmad2024proteus} that exchanges INFaaS's objective and constraint: at each load level, it deploys the highest-quality variant whose latency satisfies the SLO. INFaaS-Acc hosts a single variant across the cluster at any time. Under higher load it switches to a variant with higher latency-bounded throughput, and to a higher-quality one when capacity allows. It is not query-aware: every query is served by the active variant regardless of difficulty.
    \item \textbf{Proteus}~\cite{ahmad2024proteus} is a high-throughput serving system that adaptively selects among multiple model variants as workload changes. Unlike INFaaS-Acc, Proteus can host multiple variants concurrently across workers, drawn from the four diffusion model candidates. It scales toward faster variants under high load and toward higher-quality variants when capacity allows, and runs single-stage inference without a per-query cascade or discriminator. Queries are routed to workers uniformly at random.
    \item \textbf{DiffServe}~\cite{ahmad2025diffserveefficientlyservingtexttoimage} is a text-to-image serving system built around a model-cascading pipeline. 
    DiffServe uses an MILP planner to adapt to demand by jointly setting resource allocations and the quality threshold of the discriminator. Unlike \pjn{}, it employs a fixed pair of light and heavyweight models. In our implementation, we fix the pair to SD3.5-Turbo and SD3.5-Large, which outperformed other pairings on the query workload.
\end{itemize}

\subsection{Performance on Real-World Traces} \label{sec:real_traces}

Figure~\ref{fig:end2end_results} reports time series of demand, FID, and SLO violations, plus their averages over the trace.
Clipper-Light has the lowest quality (highest FID) and low violations because it processes all queries with the fastest model.
Clipper-Heavy uses the highest-quality model, achieving the best FID among baselines but incurring significant violations (up to 90\% at peak, 77\% on average).
INFaaS-Acc optimizes for quality and starts from the heaviest model whenever the latency budget allows. Its greedy planner reacts only to observed congestion without forecasting demand, so queues accumulate during demand ramps and produce significant SLO violation spikes (up to 26\% during ramp-up, 13.4\% on average).
Proteus dynamically tunes the resource allocation according to demand changes, selecting higher-quality variants under low load and shifting toward faster variants under high load. Unlike INFaaS-Acc, Proteus forecasts future demand using an EWMA estimator, which avoids the spikes seen during ramp-up. However, as it is not query-aware, queries are routed to workers uniformly at random regardless of difficulty, so easy queries could be over-served by heavy variants while hard queries could be under-served by light variants, thus its quality improvement is limited and it suffers from high SLO violations (up to 12\% during the peak, 7\% on average).
DiffServe offers a better response quality and a lower SLO violation ratio due to its dynamic resource allocation and query-awareness. It routes queries to the heavyweight model given the decision made by the discriminator, thus improving the quality over Proteus by up to 12\% in the timeseries and 4\% on average.

\pjn{} jointly optimizes all three dimensions of the coupled space: query-aware routing, adaptive model-pair selection, and resource allocation. As demand rises, it shifts traffic toward faster cascades and tightens heavyweight usage to protect latency; as demand decreases, it reallocates GPUs to higher-quality cascades and relaxes thresholds to recover quality. Consequently, \pjn{} improves quality by 10\%-35\% over baselines while keeping violations low across the trace. At the demand peak, \pjn{} achieves better FID than all baselines except Clipper-Heavy, which suffers significantly higher SLO violations. On average, \pjn{} reduces FID by 7\%-23\% relative to baselines (excluding Clipper-Heavy) and reduces the SLO violation ratio by 2.7$\times$, 4.2$\times$, 8$\times$, and 45$\times$ compared to DiffServe, Proteus, INFaaS-Acc, and Clipper-Heavy. Each baseline fixes at least one dimension: DiffServe fixes the model pair, Proteus and INFaaS-Acc ignore per-query routing, and Clipper fixes both.

\begin{figure}[t]
    \centering
    \includegraphics[width=0.505\linewidth]{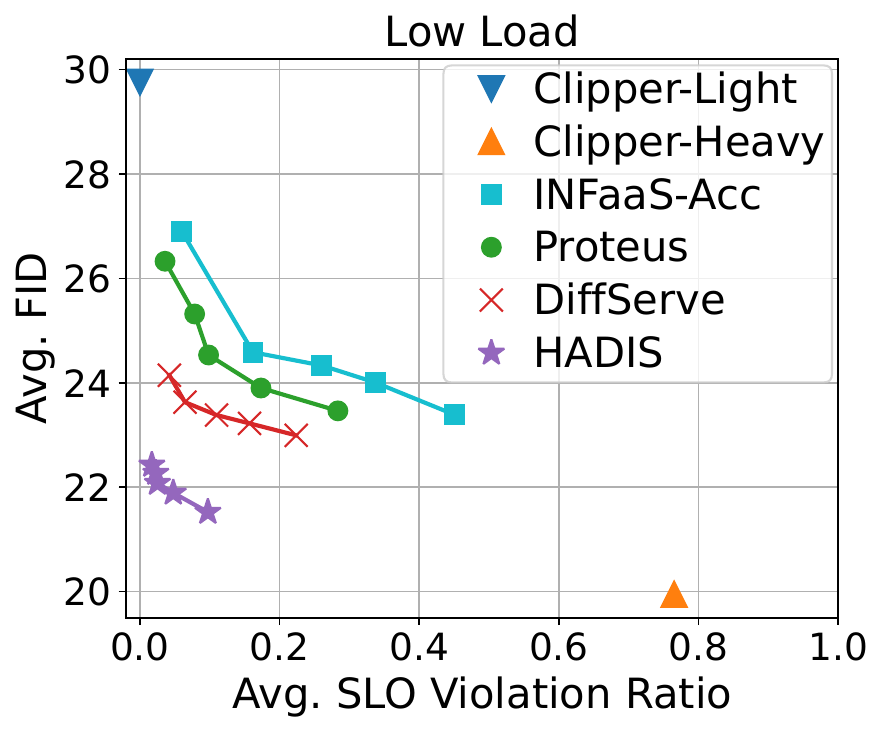}
    \hspace{-8pt}
    \includegraphics[width=0.505\linewidth]{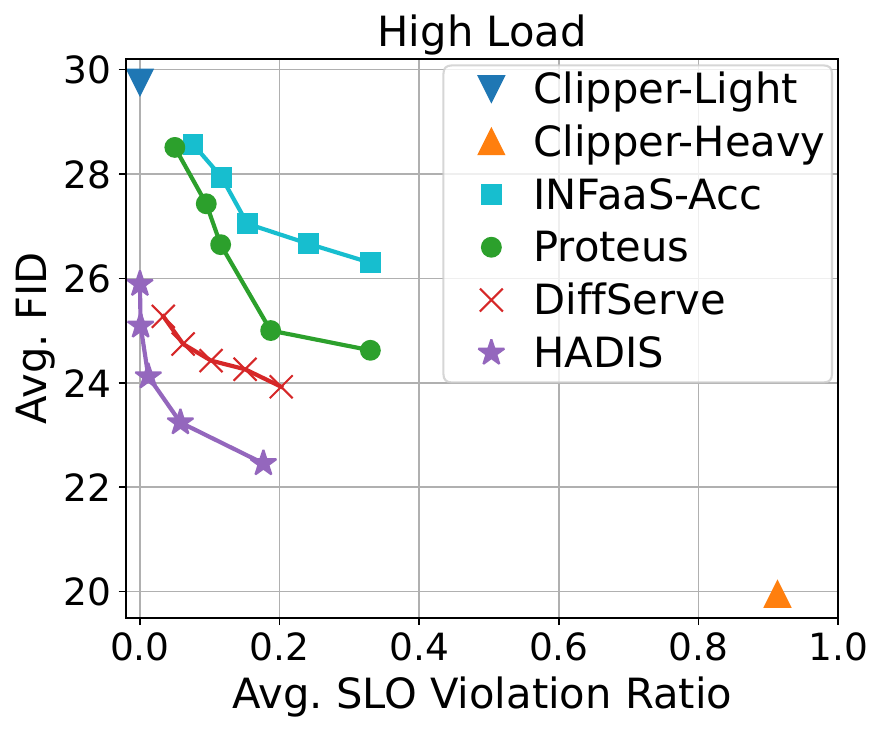}
    \vspace{-0.5cm}
    \caption{Performance comparison on static workloads. \pjn{} offers a Pareto-optimal trade-off between FID and SLO violations (lower-left curve).}
    \label{fig:tradeoff}
\end{figure}

\subsection{Performance on Static Traces} \label{sec:static_traces}
To characterize the trade-off between quality and SLO violation across systems, we evaluate \pjn{} and baselines on static workloads at low and high loads (Figure~\ref{fig:tradeoff}). At each load, we sweep an over-provisioning knob per system to generate a trade-off curve. For \pjn{}, Proteus, and DiffServe, we vary the over-provisioning factor on demand estimation (\S\ref{sec:resource_management}), which scales the predicted load fed to the planner: a larger factor reserves headroom and reduces violations at some quality cost, while a smaller factor frees capacity for heavier variants at the risk of violations. INFaaS-Acc has no demand forecasting, so we instead tighten or relax its SLO buffer to produce the corresponding trade-off. Clipper-Light and Clipper-Heavy are static and appear as single points.

\pjn{}'s curve lies strictly below-left of every baseline's at both loads, indicating Pareto-dominance: for any chosen SLO target, \pjn{} achieves lower FID; for any chosen FID target, \pjn{} achieves fewer violations. The residual SLO violations \pjn{} incurs at low load (seen in \S\ref{sec:real_traces} and \S\ref{sec:synthetic_traces}) represent a chosen operating point on this curve, and can be reduced toward zero by setting a larger over-provisioning factor at a small quality cost.

\begin{figure}[t]
    \centering
    \includegraphics[width=0.377\linewidth]{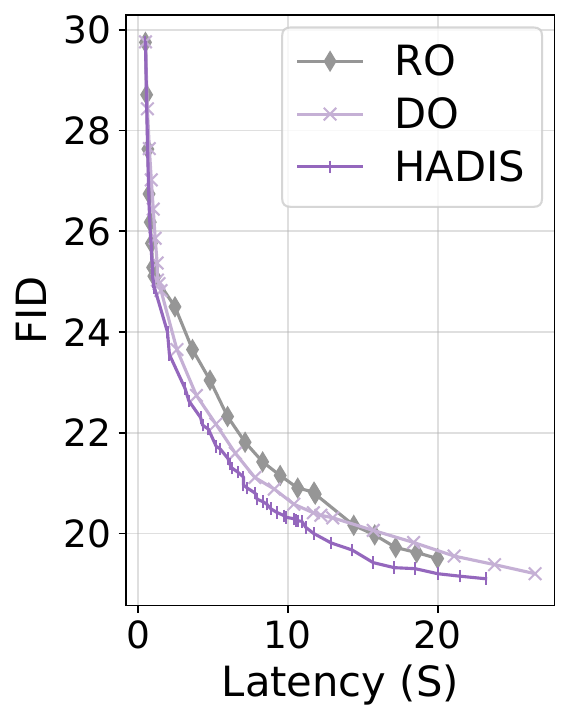}
    \hspace{-6.5pt}
    \includegraphics[width=0.63\linewidth]{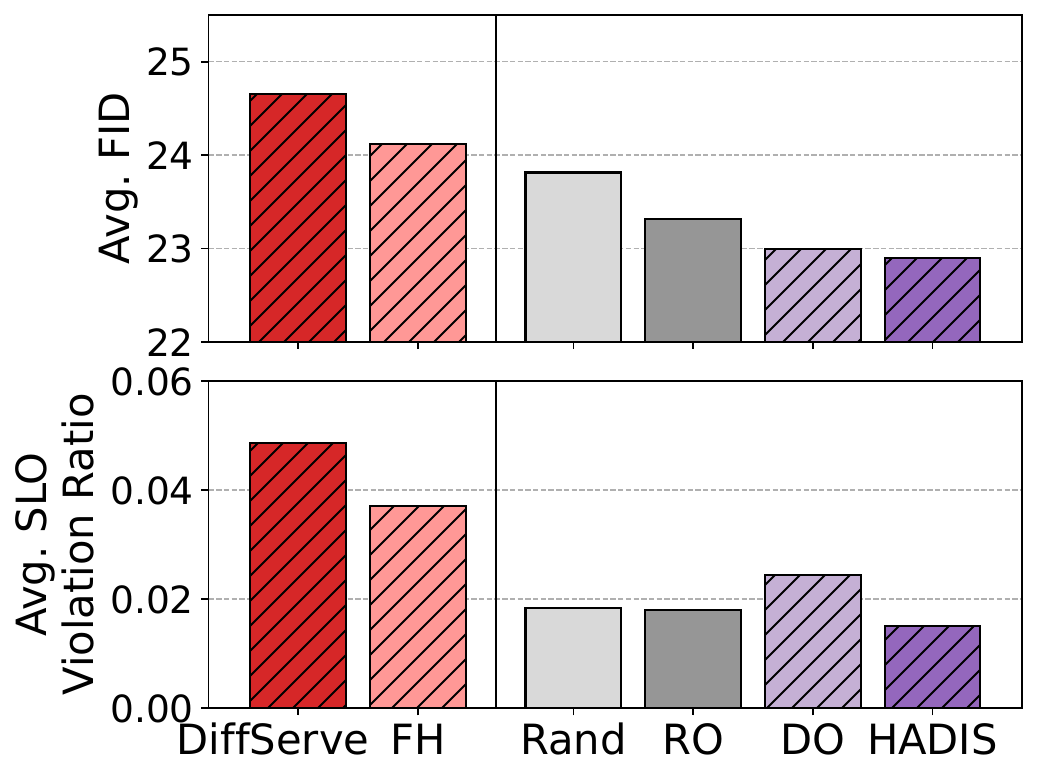}
    \caption{Effectiveness of hybrid routing and dynamic model-pair selection. \underline{(Left)} Pareto frontiers in latency-FID space; \pjn{} dominates Router-only (RO) and Discriminator-only (DO). \underline{(Right)} Average FID (\textit{top}) and SLO violation ratio (\textit{bottom}) on the real-world trace. Systems left of the divider fix the model pair; systems to the right select it at runtime. Hatched bars attribute \pjn{}'s gain over DiffServe: DiffServe and DO route with the discriminator alone, Fixed-pair Hybrid (FH) and \pjn{} with both components.}
    \label{fig:ablation_router}
\end{figure}

\subsection{Evaluation of Routing and Model-Pair Selection} \label{sec:ablation_routing}
Figure~\ref{fig:ablation_router} ablates the two mechanisms \pjn{} adds to a cascade. Three variants share \pjn{}'s runtime model-pair selection but change the routing: \textbf{Random} assigns random hardness and quality scores, \textbf{Router-only} (RO) routes by the rule-based router without discriminator re-routing, and \textbf{Discriminator-only} (DO) routes by the discriminator without the router. Two further systems hold the pair fixed at SD3.5-Turbo and SD3.5-Large: \textbf{DiffServe}, which routes with the discriminator alone, and \textbf{Fixed-pair Hybrid} (FH), which uses hybrid routing. All six run on the same real-world trace under identical resources.

Figure~\ref{fig:ablation_router}-\textit{left} shows the Pareto frontiers of RO, DO, and \pjn{} in latency-FID space across cascade configurations. DO outperforms RO at mid-latency because the discriminator classifies queries more accurately than the rule-based router. RO outperforms DO at high latency, where DO pays a slow first-stage cost on every query that RO bypasses for predicted-hard ones. \pjn{}'s hybrid design combines both strengths and traces the lowest FID throughout.

Figure~\ref{fig:ablation_router}-\textit{right} shows that \pjn{} achieves the lowest average FID and SLO violation ratio across all variants that select the pair at runtime.
Random and Router-only suffer worse FID at similar SLO violation levels since many \textit{hard} queries are served by lightweight models due to misclassification. Discriminator-only improves FID but incurs higher violations, as it forces every query through the light model, paying extra latency. In contrast, \pjn{} adopts hybrid routing, skipping light stage for obviously hard queries and catching false ``easy'' cases with discriminator. Thus, it preserves low latency and achieves higher response quality.

The hatched bars separate the two sources of \pjn{}'s gain over DiffServe. At a fixed model pair, FH reduces the average SLO violation ratio from 4.8\% to 3.7\% and FID from 24.7 to 24.1, isolating the benefit of hybrid routing. With discriminator-only routing, DO achieves 2.4\% and 23, isolating the benefit of runtime model-pair selection. \pjn{} combines both mechanisms and improves on either alone.

We evaluate the router via end-to-end system metrics (FID and SLO violations) rather than per-query accuracy rates because text-to-image quality is measured by FID, a distributional metric over query populations, not a per-sample correctness label. No ground-truth ``easy'' vs.\ ``hard'' classification exists for individual prompts.

\begin{figure}[t]
    \centering
    \includegraphics[width=0.94\linewidth]{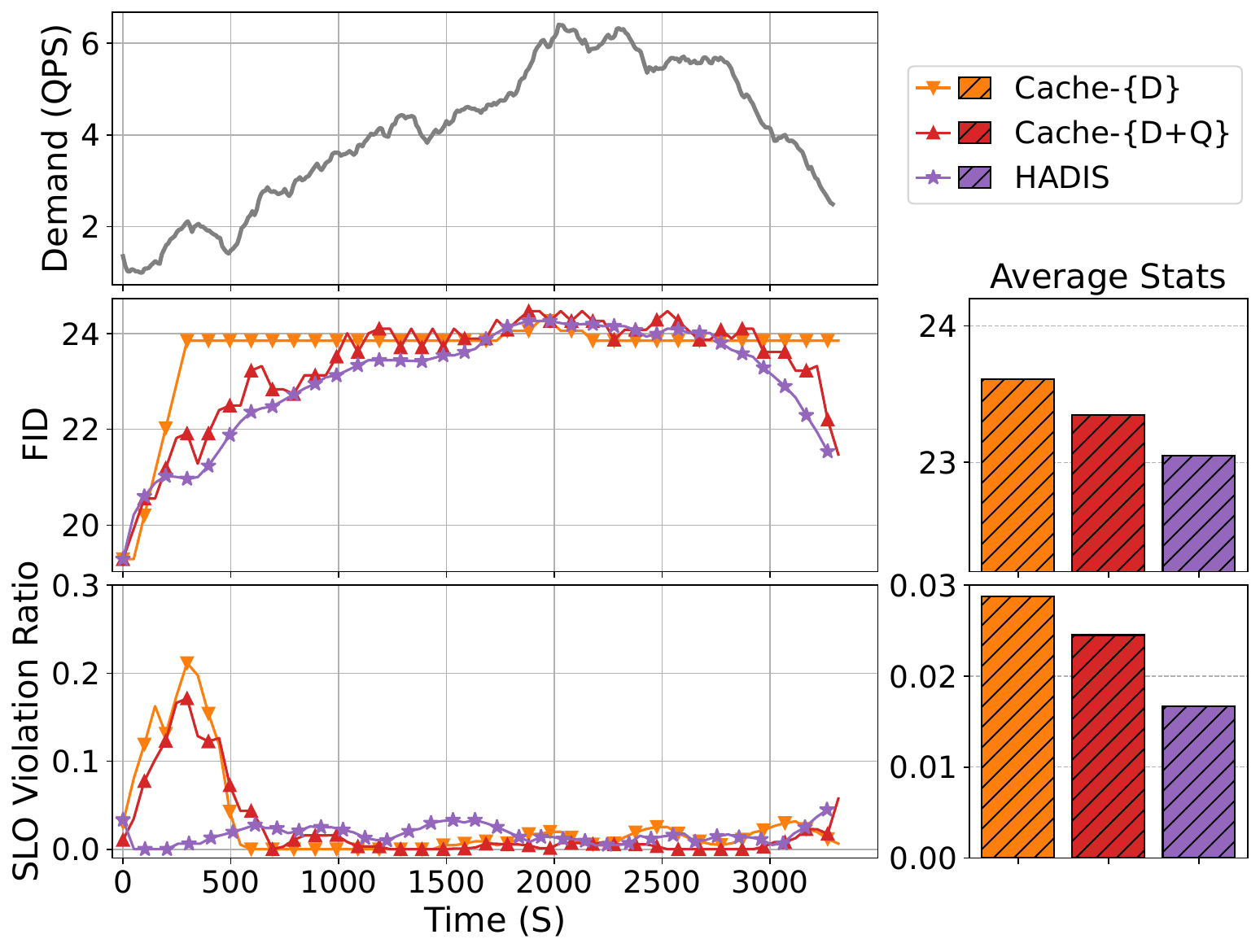}
    \vspace{-0.3cm}
    \caption{Comparison of resource allocation strategies. \pjn{} applies MILP. Cache-\{D\} caches plans using demands as key; Cache-\{D+Q\} caches plans using demands and queue length as key. Cached policies lag during load shifts, showing transient quality drops and violation spikes, whereas \pjn{} adapts in real time, yielding lower violations with comparable or better FID.}
    \label{fig:ablation_milp}
\end{figure}

\subsection{Evaluation of Resource Allocation} \label{sec:ablation_milp}
We ablate the resource planner by implementing two variants, \textbf{Cache-\{D+Q\}} and \textbf{Cache-\{D\}}. Unlike \pjn{} which solves an online MILP to seek optimal resource allocation plans at runtime, these cached planners reuse previously computed plans on a cache hit. Cache-\{D\} indexes plans by a discretized \emph{range} of system demand, whereas Cache-\{D+Q\} uses a two-dimensional key over \emph{range} of system demand and \emph{range} of worker queue length. 
At runtime, we quantize observed demand (and queue length for Cache-\{D+Q\}) to their ranges and probe the cache. If the measurements fall within a keyed range with an associated plan, we declare a cache hit and apply that plan. If not, we solve the MILP and insert the new plan under the corresponding range key. 

Figure~\ref{fig:ablation_milp} shows the timeseries results and average statistics. Cache-\{D\} performs the worst, with early spikes in FID and SLO violations because it lags behind load shifts and is agnostic about real-time system workload. Cache-\{D+Q\} is more responsive by incorporating queue length into the key to capture congestion. We empirically observe that both caching variants exhibit a startup spike in SLO violations since their initial plans are either biased or misaligned with the actual demands, but violations stabilize once the workload settles. 
In contrast, \pjn{} with online MILP adapts plans in real time, rebalancing workers and thresholds as demand changes. It prevents prolonged queue growth, keeps violations low, and maintains comparable or better FID, avoiding under- or over-provision when workload fluctuates.
Moving from Cache-\{D\} to Cache-\{D+Q\} improves both metrics, suggesting that richer keys help but do not substitute for online re-optimization.

\begin{figure}[t]
    \centering
    \includegraphics[width=0.94\linewidth]{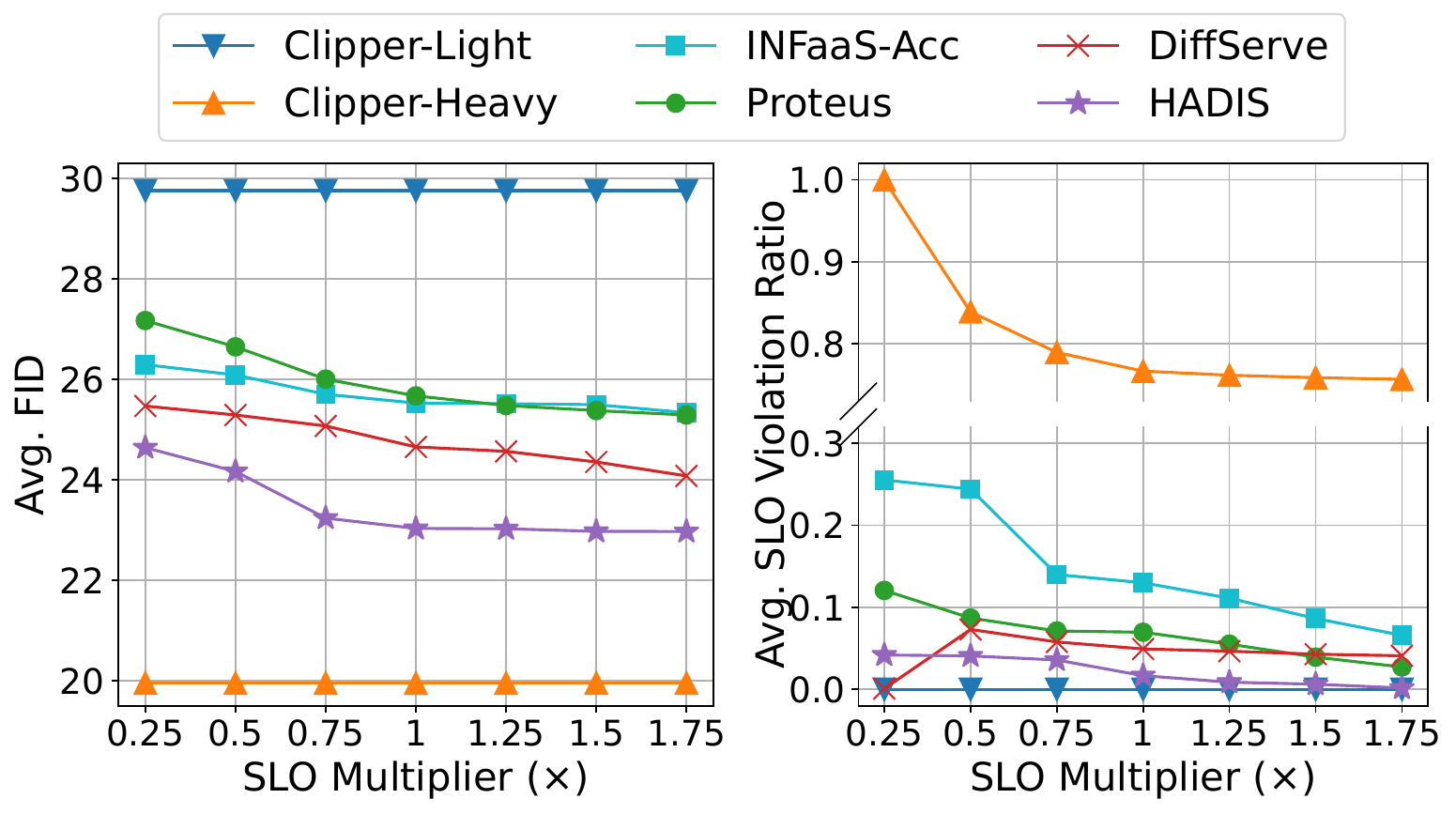}
    \vspace{-0.3cm}
    \caption{Sensitivity to latency SLO on \pjn{} and baselines. \pjn{} guarantees low SLO violations and high quality over varying SLO values across all methods. }
    \label{fig:ablation_slo}
\end{figure}

\begin{figure*}[t]
  \centering
  \begin{subfigure}{0.32\linewidth}
    \includegraphics[width=\linewidth]{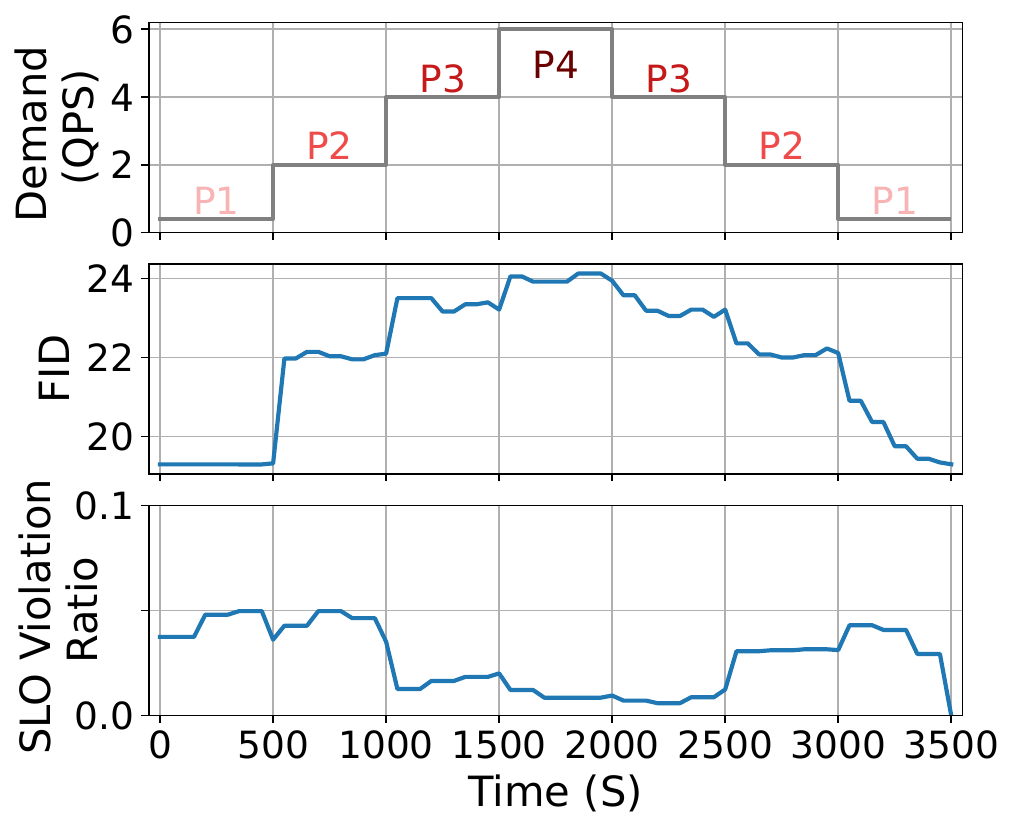}
    \vspace{-0.6cm}
    \caption{Demand trace and system performance}
    \label{fig:end2end_step_results}
  \end{subfigure}\hfill
  \begin{subfigure}{0.67\linewidth}
    \includegraphics[width=\linewidth]{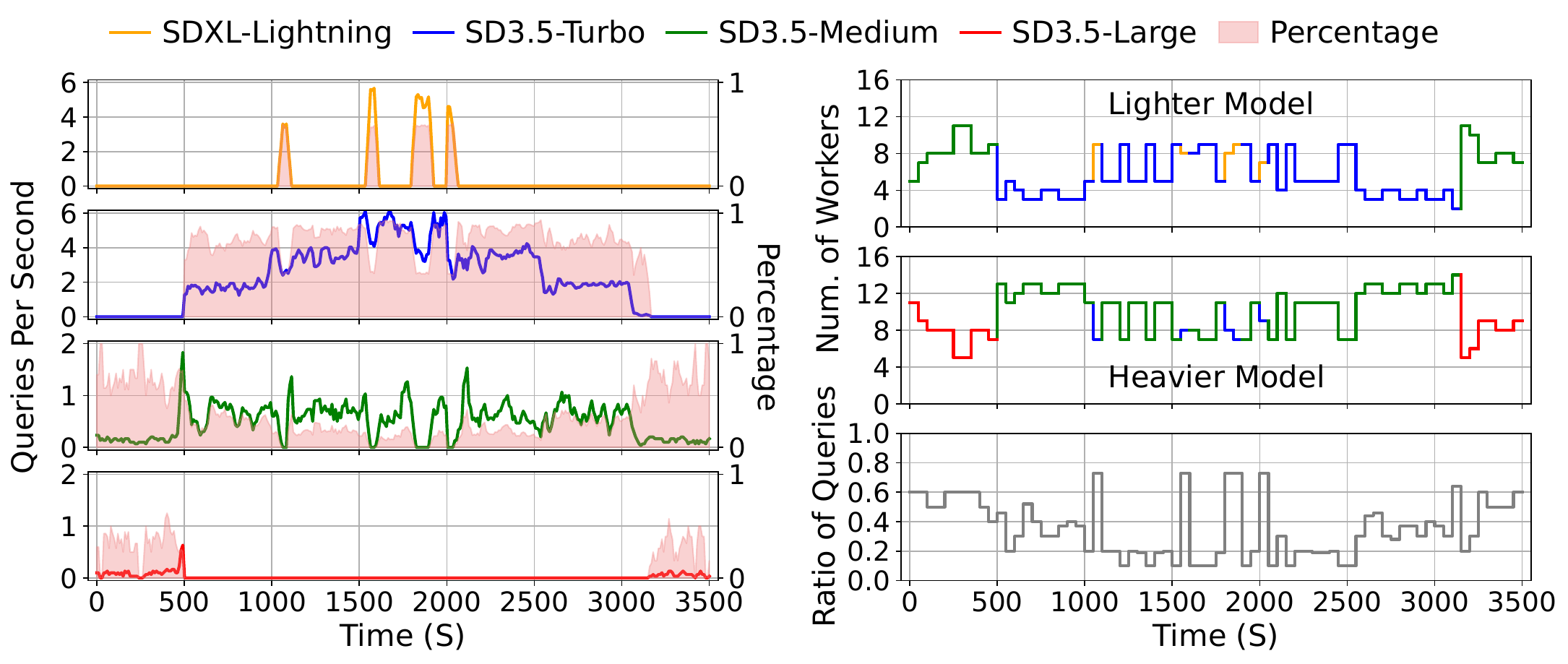}
    \vspace{-0.6cm}
    \caption{Per-model queries per second processed (\textit{left}) and dynamic resource allocation (\textit{right})} 
    \label{fig:end2end_step_resource}
  \end{subfigure}
  \vspace{-0.3cm}
  \caption{Performance of \pjn{} on synthetic traces.
  \underline{(a)} Time-series across four demand phases (P1-P4). \textbf{Top: } demand QPS. \textbf{Middle: } FID rises when demand increases and recovers as demand decreases. \textbf{Bottom: } SLO violation ratio stays low overall, rising slightly during P1/2 when heavier models dominate. 
  \underline{(b)} \textbf{Left: } queries are shifted toward faster variants under higher load and back toward higher-quality variants as load decreases. \textbf{Right: } \textit{worker allocation} is coordinated with \textit{the ratio of queries routed to heavyweight models} to maintain low SLO violations while maximizing quality.}
  \label{fig:end2end_analysis}
\end{figure*}

\subsection{Performance on Synthetic Traces} \label{sec:synthetic_traces}
We evaluate \pjn{} on a controlled synthetic workload to illustrate how it adapts routing, model selection, and resource allocation over time (Figure~\ref{fig:end2end_analysis}). The trace replays four piecewise-constant load levels (P1--P4) arranged as rising, peak, and falling phases, mimicking bursty traffic.

Figure~\ref{fig:end2end_step_results} shows that as demand climbs (P3/P4), FID increases modestly as \pjn{} trades quality for throughput; when demand falls (P1/P2), quality recovers. SLO violations remain low throughout (1--5\%). Violations are slightly higher during P1/P2 because the system uses SD3.5-Large and SD3.5-Medium for higher quality, which have longer execution latency.

Figure~\ref{fig:end2end_step_resource}-\textit{left} shows the traffic mix shifting toward faster models (SDXL-Lightning, SD3.5-Turbo) during P3/P4 and back toward higher-quality models during P1, as the MILP selects configurations from the profiled table that meet the SLO under each load level. Figure~\ref{fig:end2end_step_resource}-\textit{right} shows that worker allocation is coordinated with routing: more workers are assigned to heavyweight models when more queries are routed to them. In P3/P4, \pjn{} periodically uses SDXL-Lightning to drain queued queries accumulated by heavier models, preventing timeouts while maintaining high quality.

\subsection{Sensitivity to Latency SLO} \label{sec:ablation_slo}
We vary the SLO multiplier from 0.25$\times$ to 1.75$\times$ and run \pjn{} and all baselines on the real-world trace. This range is chosen relative to the model runtimes, so that it spans deadlines below the heavyweight model's runtime at 0.25$\times$ and deadlines well above every model's runtime at 1.75$\times$. Figure~\ref{fig:ablation_slo} reports the average FID and SLO violation ratio.

As the SLO relaxes, all methods improve on both metrics, but \pjn{} outperforms baselines across the entire range. Under tight budgets (0.25-0.75$\times$), \pjn{} sustains low violations while achieving lower FID than Proteus, INFaaS-Acc, and DiffServe (up to 11\%, 7\%, and 7\%). Clipper-Light maintains $\sim$0 violations but delivers the worst FID, whereas Clipper-Heavy attains the best FID yet exhibits very high SLO violation ratios even with higher SLOs.
The SLO violation ratio of DiffServe is nearly zero at the 0.25$\times$ multiplier because the latency SLO (15s) is below the runtime of the heavyweight model ($\sim$27s), making the MILP allocate all workers to lightweight models.
Although \pjn{} could also select a lighter cascade, sacrificing FID for near-zero SLO violations, this would violate its objective of maximizing quality under SLO constraints. Thus, the planner avoids such degenerate allocations.
For SLO multipliers above $1\times$, all methods become demand-bound rather than deadline-bound, where performance is limited by the offered QPS. Results show that \pjn{} guarantees low SLO violations and high quality over a broad range of SLOs.

\begin{figure}[t]
    \centering
    \includegraphics[width=0.92\linewidth]{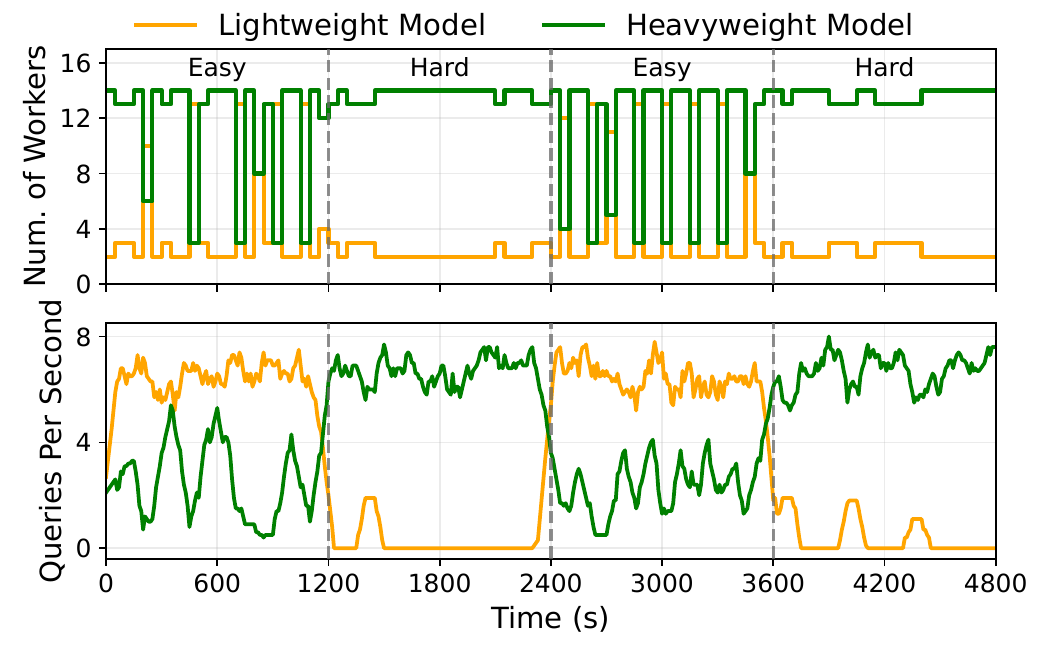}
    \vspace{-0.3cm}
    \caption{Sensitivity to Query Hardness. \pjn{} adaptively bypasses lightweight inference and reallocates workers according to query complexity.}
    \label{fig:ablation_prompt_difficulty}
\end{figure}

\subsection{Sensitivity to Hardness of Queries} \label{sec:eval_hardness}
We study how sensitive \pjn{} is to shifts in the hardness distribution of incoming queries. We isolate the effect of hardness from demand by fixing the query demand and constructing a four-phase trace: the first and third phases contain only \textit{Easy} queries, and the second and fourth contain only \textit{Hard} queries. Queries are categorized using weighted scores over textual features.
As shown in Figure~\ref{fig:ablation_prompt_difficulty}, during Easy phases, the system settles into a relatively dynamic phase where the lightweight model serves the majority of traffic. 
At each transition to a Hard phase, the router bypasses lightweight inference for a larger fraction of queries, sharply increasing the heavy-class intake. The heavy queue rises, triggering workers assigned to the heavyweight model. 
\pjn{} thus adaptively concentrates resources on heavyweight models precisely when hard queries predominate, highlighting the value of our query-side bypass design.

\begin{figure}[t]
    \centering
    \includegraphics[width=0.33\linewidth]{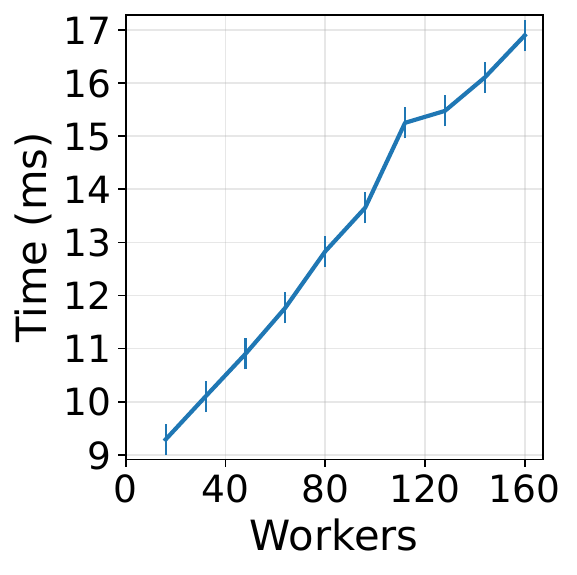}
    \hspace{-6pt}
    \includegraphics[width=0.34\linewidth]{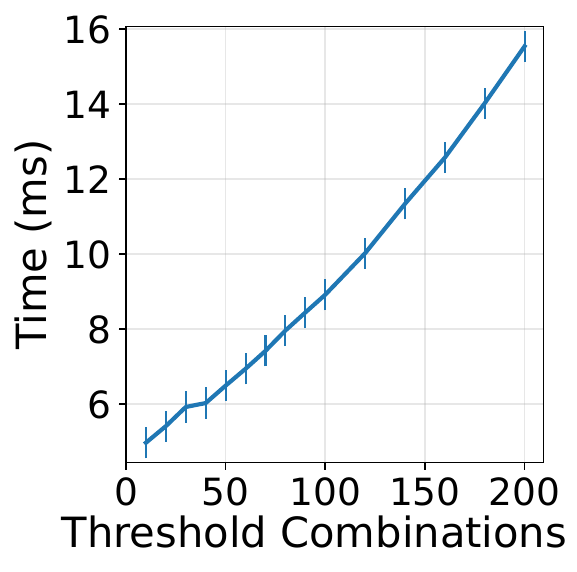}
    \hspace{-6pt}
    \includegraphics[width=0.325\linewidth]{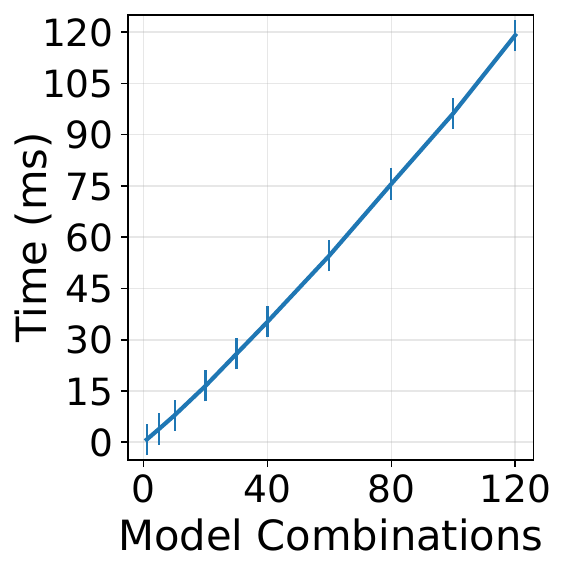}
    \caption{Scalability of MILP with respect to workers, threshold combinations, and model combinations. Solve time scales roughly linearly with each dimension and stays in the millisecond range across all tested values.}
    \label{fig:overhead_milp}
\end{figure}

\subsection{Overhead Analysis} \label{sec:overhead_analysis} 
\pjn{} introduces four sources of overhead: 
(1) a query router that performs rule-based feature extraction and a score computation (5ms per query), (2) a CLIP-based discriminator that computes quality score for generated images (7ms per query on an L40S GPU), (3) an MILP-based resource allocator whose measured average runtime is $\sim$30ms per planning cycle, and (4) a model transition when a worker changes its assigned model.
The aggregated overhead of router and discriminator is $\sim$12ms, which is negligible relative to diffusion inference of SDXL-Lightning (2.4\%), SD3.5-Turbo (0.9\%), SD3.5-Medium (0.01\%), and SD3.5-Large (<0.01\%). 
Moreover, since the MILP solver runs periodically and applies updates asynchronously, this overhead is off the critical path and does not affect individual queries. 
Figure~\ref{fig:overhead_milp} shows that solve time scales roughly linearly with the number of workers, threshold combinations, and model combinations, and stays within the millisecond range across all tested values, well below diffusion inference latency. The MILP overhead therefore remains negligible at deployments substantially larger than our 16-worker testbed.
We set the planning interval to 50s for the traces evaluated here. A model transition takes at most 1.32s, under 3\% of the planning interval and the latency SLO. A worker pays it once per reconfiguration rather than per query, and only when its assignment changes.

\section{Discussion}

\mypar{Distribution shift}
\pjn{} fixes route ratios and router weights offline, so a sustained shift in the prompt distribution leaves the planner working from stale estimates.
\S\ref{sec:eval_hardness} shows that queue feedback still absorbs abrupt shifts in query hardness, since the resulting change in queue lengths drives subsequent planning cycles to correct the thresholds and the heavyweight allocation. The feedback does not refresh the ratios and weights themselves, however, and updating them from served traffic would close the gap without changing the architecture.

\mypar{Router and discriminator}
\pjn{} instantiates the router as a rule-based prompt analyzer and the discriminator as a CLIP classifier, but both are pluggable interfaces (\S\ref{sec:hybrid_routing}). A VLM judge would score quality more faithfully, and is practical as long as its scoring cost stays small relative to diffusion inference. \pjn{}'s contribution is the hybrid architecture and its joint optimization.

\mypar{Potential extensions}
The following settings fall outside the current implementation, and the hybrid cascade architecture is able to support them.
(1) \textit{Multi-turn editing.} \pjn{} treats every prompt independently, while users of text-to-image services often refine a prompt across several turns. A multi-turn router would take previous prompts and routing history as additional input.
(2) \textit{Text-to-video.} Video generation has the same latency-quality spectrum and no intermediate quality signal. Serving it needs profiled video models and a discriminator that scores temporal consistency.
(3) \textit{Mixed SLOs.} Requests may arrive with different deadlines rather than the single SLO. Admitting a range of deadlines would change the MILP formulation but not the cascade architecture.
(4) \textit{Broader model pools.} The two-model result of \S\ref{sec:background} is an empirical observation over the model pool and latency range we evaluate. Profiling and the MILP handle a larger pool without change, and evaluating a wider range of models and latencies would show how far the observation holds.
We leave these extensions to future work.
\section{Related Work}

Model serving systems aim to provide a unified abstraction to the user, hiding details of the underlying ML frameworks, data preprocessing, and performance optimization.
Systems such as SageMaker~\cite{sagemaker2020build}, Triton~\cite{Triton}, TensorFlow Serving~\cite{Tensorflow}, TorchServe~\cite{PyTorch}, and research prototypes like Nexus~\cite{shen2019nexus}, BATCH~\cite{ali2020batch}, and Clockwork~\cite{gujarati2020serving} require users to specify which models to deploy; the system handles resource management but does not reason about model variants.

\textbf{Variant-aware serving.}
More recent systems jointly optimize accuracy, latency, and cost by managing model variants. Clipper~\cite{crankshaw2017clipper}, Rafiki~\cite{wang2018rafiki}, and Cocktail~\cite{gunasekaran2022cocktail} ensemble models to improve quality. INFaaS~\cite{romero2021infaas}, Model Switching~\cite{zhang2020model}, and Sommelier~\cite{guo2022sommelier} adaptively select variants under changing load. Proteus~\cite{ahmad2024proteus} further exploits accuracy-efficiency trade-offs across variants and represents the state-of-the-art in adaptive variant selection for general inference serving. However, all these systems react to system-level load signals and serve all queries identically regardless of difficulty. When applied to diffusion models, Proteus adapts \textit{which} variants are deployed but not \textit{how} individual queries are routed. \pjn{} outperforms Proteus by 4.2$\times$ in SLO violations (\S\ref{sec:evaluation}) because per-query routing captures optimization opportunities that variant selection alone misses.

\textbf{Input-adaptive inference.}
A separate class of systems varies the computation spent on each input. SkipNet~\cite{Wang_2018_ECCV} skips layers of a CNN conditioned on the input, and early-exit designs let easy inputs terminate at intermediate layers. HELIOS~\cite{kumar2025heliosadaptivemodelearlyexit} applies early exit to LLM serving, switching among early-exit models to increase the fraction of tokens that exit early. AdaDiff~\cite{zhang2025adadiff} learns an instance-specific policy over the number of denoising steps each prompt receives.
However, all these methods require models or policies trained for dynamic execution, and that training is specific to the model being served. \pjn{} composes off-the-shelf variants into an external cascade and adapts by swapping whole models rather than execution paths within one. 

\textbf{Cascade and query-aware serving.}
Cascaded architectures (CascadeBERT~\cite{li2020cascadebert}, Tabi~\cite{wang2023tabi}, CascadeServe~\cite{kossmann2024cascadeserve}) route easy queries to lightweight models and defer harder ones to heavier models. DiffServe~\cite{ahmad2025diffserveefficientlyservingtexttoimage} extends this to diffusion models, constructing a two-model cascade and formulating resource allocation as an MILP. 
Argus~\cite{10.1145/3721462.3770765} selects one approximation option for each query to reduce generation cost and serves the query once with that option. 
NIRVANA~\cite{295597} reduces cost by caching intermediate denoising states within a single model variant, and MoDM~\cite{modm2026} reuses previously generated images with smaller models to avoid redundant generation; both are complementary to query-aware cascading. All cascade frameworks, however, fix the model pair and always begin with a lightweight pass, wasting computation on queries that are clearly hard.
\pjn{} addresses both gaps: it combines pre-generation routing with post-generation quality verification that catches incorrect routing decisions, and jointly optimizes the model pair, thresholds, and GPU allocation at runtime. 

\section{Conclusion}
\pjn{} is a diffusion model serving system that resolves a fundamental tension in cascade serving: all queries must pass through the lightweight stage before quality can be assessed, yet predicting difficulty before generation remains unreliable. The hybrid cascade architecture combines a pre-generation router with a post-generation discriminator, addressing complementary failure modes that neither component handles alone. We show that two-model cascades capture the Pareto frontier of deeper cascades, reducing the coupled optimization space (model pair, thresholds, GPU allocation) to a compact configuration table that an online MILP re-solves in milliseconds. On a 16-GPU cluster with real-world traces, \pjn{} improves response quality by up to 35\% and reduces SLO violations by 2.7--45$\times$ compared to baselines.

\begin{acks}
This material is based upon work supported by the National Science Foundation under Grant No. CNS-2312396, CNS-2338512, IIS-2435822, and CCF-2449995.
Any opinions, findings, conclusions, or recommendations expressed in this material are those of the author(s) and do not necessarily reflect the views of the National Science Foundation. 
We also thank the anonymous reviewers and our shepherd for their valuable feedback that helped improve this paper.
\end{acks}

\bibliographystyle{ACM-Reference-Format}
\bibliography{ref}

\end{document}